\documentclass[lettersize,journal]{IEEEtran}
\usepackage{amsmath,amsfonts}
\usepackage{graphicx,multirow,booktabs,diagbox}
\usepackage{tablefootnote}
\usepackage[colorlinks=true, linkcolor=black, citecolor=black, urlcolor=black]{hyperref}
\usepackage[table]{xcolor}
\usepackage{makecell}
\usepackage{tikz, pgfplots}

\ifCLASSOPTIONcompsoc
    \usepackage[caption=false, font=normalsize, labelfont=sf, textfont=sf]{subfig}
\else
\usepackage[caption=false, font=footnotesize]{subfig}
\fi
\newcommand{\tabincell}[2]{\begin{tabular}{@{}#1@{}}#2\end{tabular}}

\definecolor{myBlue}{HTML}{4477AA}
\definecolor{myRed}{HTML}{BB5566}
\definecolor{myGreen}{HTML}{228833}
\definecolor{myYellow}{HTML}{DDAA33}

\def\x{{\mathbf{x}}}

\def\h{{\mathbf{h}}}
\def\A{{\mathbf{A}}}

\begin{document}

\title{Leveraging ASR Pretrained Conformers for\\ Speaker Verification through \\Transfer Learning and Knowledge Distillation}

\author{Danwei~Cai,~\IEEEmembership{Student Member,~IEEE}
        and~Ming~Li,~\IEEEmembership{Senior Member,~IEEE}%
\thanks{D. Cai, and M. Li are with the Department of Electrical and Computer Engineering, Duke University, Durham, NC, 27705, USA, e-mail: \{danwei.cai, ming.li369\}@duke.edu.}%
\thanks{M. Li is also with the Suzhou Municipal Key Laboratory of Multimodal Intelligent Systems, Data Science Research Center, Duke Kunshan University, Kunshan, China.}
\thanks{Corresponding author: Ming Li.}%
}


\maketitle

\begin{abstract}
This paper focuses on the application of Conformers in speaker verification.
Conformers, initially designed for Automatic Speech Recognition (ASR), excel at modeling both local and global contexts within speech signals effectively.
Previous research has established that ASR and speaker verification tasks can naturally complement each other.
Building on this synergistic relationship, this study introduces three strategies for leveraging ASR-pretrained Conformers in speaker verification:
(1) Transfer learning: We use a pretrained ASR Conformer encoder to initialize the speaker embedding network, thereby enhancing model generalization and mitigating the risk of overfitting.
(2) Knowledge distillation: We distill the complex capabilities of an ASR Conformer into a  speaker verification model. This not only allows for flexibility in the student mode's network architecture but also incorporates frame-level ASR distillation loss as an auxiliary task to reinforce speaker verification.
(3) Parameter-efficient transfer learning with speaker adaptation: A lightweight  \textcolor{black}{speaker adaptation module} is proposed to convert ASR-derived features into speaker-specific embeddings, without altering the core architecture of the original ASR Conformer. This strategy facilitates the concurrent execution of ASR and speaker verification tasks within a singular model.
Experiments were conducted on VoxCeleb datasets.
The results are compelling: models employing ASR pretraining and knowledge distillation significantly outperform standard Conformers.
Specifically, the best model using the ASR pretraining method achieved a 0.43\% equal error rate (EER) on the VoxCeleb1-O test trial, while the knowledge distillation approach yielded a 0.38\% EER.
Furthermore, by adding a mere 4.92 million parameters to a 130.94 million-parameter ASR Conformer encoder, the speaker adaptation approach achieved a 0.45\% EER, enabling parallel speech recognition and speaker verification within a single ASR Conformer encoder.
Overall, our techniques successfully transfer rich ASR knowledge to advanced speaker modeling.
\end{abstract}


\section{Introduction}
\IEEEPARstart{S}{peaker} verification, which analyzes speech signals to verify the speaker's identity, has many applications, from voice assistants to security systems.
Over the past five years, the performance of speaker verification systems has improved remarkably due to the application of deep neural networks (DNN) \cite{snyder_x-vectors:_2018,cai_--fly_2020}.
Numerous innovations have been introduced in network architecture \cite{cai_exploring_2018,okabe_attentive_2018,zhou_resnext_2021,desplanques_ecapa-tdnn_2020}, training objectives \cite{deng_arcface_2019, 9023039, chung2020in}, and training strategies \cite{garcia-romero_magneto_2020, 9414600} specifically tailored to speaker verification models.

Prevalent network architectures in speaker verification systems are convolutional neural networks (CNNs) and time-delay neural networks (TDNNs).
The key strength of CNNs and TDNNs lies in their ability to model local feature patterns effectively, which is crucial in identifying speaker-specific vocal traits.
These networks have been further advanced through variants of CNN and TDNN that incorporate residual connections \cite{he_deep_2016}, squeeze and excitation operations \cite{Hu_2018_CVPR, desplanques_ecapa-tdnn_2020}, Res2Net blocks \cite{8821313, zhou_resnext_2021, desplanques_ecapa-tdnn_2020}, and ResNeXt blocks \cite{Xie_2017_CVPR, zhou_resnext_2021}.
These modifications have significantly improved speaker verification performance.

Despite their successful applications, TDNNs, CNNs, and their variants face limitations in extracting long-range global context, especially without deep layers.
As an alternative, Transformers, with their multi-head attention mechanism, have demonstrated a more robust ability to capture global context with less fine-grained local patterns \cite{vaswani2017attention}.
To bridge this gap, Conformer combines the convolution module with Transformer to effectively capture local and global contextual information, leading to promising results in end-to-end automatic speech recognition (ASR) \cite{gulati_conformer_2020}.
Recently, Zhang \textit{et al.} introduced multi-scale feature aggregation Conformer (MFA-Conformer) for speaker verification \cite{zhang_mfa-conformer_2022}.
MFA-Conformer concatenates frame-level outputs from all Conformer blocks to enhance speaker trait extraction in speaker verification.
Liao \textit{et al.} equipped the Conformer encoder with length-scaled attention and sharpness-aware minimization training for speaker verification \cite{10095433}.
\textcolor{black}{However, despite their strengths, Conformers are susceptible to overfitting, particularly when faced with limited data or when employing large model parameters. This challenge is acute in speaker verification, where the diversity and amount of training data may be constrained \cite{zhang_mfa-conformer_2022, 10096659}.}

The Conformer model's ability to capture both local and global contexts is leveraged in ASR and speaker verification.
ASR focuses on recognizing the linguistic content of the speech, with a higher emphasis on frame-level details.
In contrast, speaker verification targets identifying speaker-specific traits derived from the speech, centering on utterance-level context.
Despite these differences, the two tasks can complement each other.
For instance, the frame-level phoneme modeling undertaken in ASR could support speaker verification by aiding the detection of unique speaker-specific articulation patterns.
Prior studies provide evidence of this synergy, showing that phoneme modeling improves speaker verification in speaker embedding networks \cite{zhou_cnn_2019} as well as the i-vector statistical model \cite{li2016generalized,6853887}.

In light of the above, our research aims at leveraging ASR Conformers for speaker verification in three distinct ways.
This builds upon our prior research on transfer learning using a pretrained ASR Conformer, which forms our first proposed method in this paper \cite{10096659}.
\textcolor{black}{The technique involves initializing the speaker embedding network with a Conformer pretrained on a large-scale ASR dataset.
This approach addresses the tendency of Conformers to overfit with limited data \cite{zhang_mfa-conformer_2022, 10096659} by leveraging a model pretrained on extensive ASR data.
The pretrained ASR Conformer, which learns rich features from a large ASR dataset, reduces the data requirements for the speaker verification task and enhance the model's generalization ability.}
Experimental results indicate that our ASR-pretrained method outperforms alternatives across various model sizes.
Notably, the best system with ASR pretraining achieved an EER of 0.48\% on the VoxCeleb 1-O trials, marking a 50\% relative improvement compared to its counterpart without ASR pretraining.

Second, we propose using knowledge distillation \cite{hinton_distilling_2015} to transfer knowledge from the ASR task to the speaker verification task.
One challenge with straightforward transfer learning is its inherent constraint on network architecture.
When using a pretrained ASR Conformer for speaker verification, the speaker model is often constrained to adopt the same network architecture as the pretrained ASR model.
To overcome this limitation, we use knowledge distillation.
In this process, a student model, a simpler neural network, is trained to mimic the behavior of the more complex, pretrained teacher ASR Conformer.
Rather than directly replicating weights and structure, knowledge distillation transfers the functional knowledge from the teacher to the student model.
This not only retains the flexibility of network architecture for the speaker verification model but also harnesses the rich information in the pretrained ASR Conformers.
Furthermore, our tailored knowledge distillation procedure, bridging ASR to speaker verification, integrates phoneme recognition as an auxiliary task.
This alignment reinforces the synergy between ASR and speaker verification tasks, ensuring the speaker verification model captures the nuanced phonetic differences recognized by the ASR Conformer.
Experimental results prove the efficacy of our method: it consistently improves speaker verification performance over the baseline method across various architectures and frequently surpasses the ASR-pretrained approach.

Finally, we propose an adaptation mechanism to unify the tasks of ASR and speaker verification within a single Conformer model.
The motivation for this approach lies in tackling the inherent inefficiency of maintaining separate models for ASR and speaker verification tasks.
Such a unified Conformer has diverse applications.
For example, our unified model streamlines the process in scenarios where ASR and speaker verification are sequentially needed, such as voice assistants authenticating a user and then transcribing their commands.
To achieve this goal, we introduce the speaker adaptation method to transform the features learned from the ASR task into those suitable for speaker verification without changing the inputs and outputs of the ASR Conformer.
The viability of this approach is supported by the speaker information preserved in the layer outputs of the ASR Conformer encoder. 
Our exploratory linear probe experiments indicate that the lower layers of the ASR Conformer retain more speaker information than the upper layers.
This speaker adaptation approach, therefore, represents a resource-efficient strategy that allows for the simultaneous and efficient execution of both ASR and speaker verification tasks using a single Conformer.
Experiments demonstrate that incorporating a \textcolor{black}{speaker adaptation module} (4.92 million parameters) into a pretrained ASR Conformer encoder (130.94 million parameters) allows for parallel execution of speech recognition and speaker verification, achieving an EER of 0.45\%.

\section{Related Works}
\subsection{Pretrained models for speaker verification}

Several studies have explored the application of self-supervised pretrained Transformers for speaker verification tasks.
Fan \textit{et al.} \cite{fan_exploring_2021}, and Vaessen \textit{et al.} \cite{vaessen_fine-tuning_2022} adopted a direct fine-tuning approach on the pretrained model by incorporating an additional pooling layer on top of the model's output.
However, this method did not surpass the performance of CNN- or TDNN-based speaker verification models, which typically have fewer parameters than the pretrained Transformer.
Novoselov \textit{et al.} \cite{novoselov_robust_2022} fine-tuned wav2vec 2.0 by integrating two simple TDNN layers and a statistic pooling layer. Their findings suggested that utilizing the entire deep pretrained encoder architecture was unnecessary, as earlier layers potentially provided more speaker information.

Another prevalent method replaces the handcrafted feature with the pretrained frame-level feature to train TDNN- or CNN-based speaker embedding networks \cite{chen_large-scale_2022,9814838}.
This approach, employing a layer-wise weighted average to aggregate features from different Transformer layers, has improved performance over models using handcrafted spectral features.
However, this comes at the cost of using a large number of pretrained parameters alongside a full TDNN- or CNN-based speaker embedding network.
Expanding on the concept of layer-wise weighted average as a feature aggregation method, Peng \textit{et al.} \cite{peng_attention-based_2022} proposed multi-head factorized attentive pooling, which can be viewed as a fusion of layer-wise weighted average and multi-head attentive pooling.

In this paper, instead of self-supervised pretrained Transformers, an ASR-pretrained Conformer is used as the network backbone for the speaker embedding network since there are already many large-scale publicly open ASR datasets available.
We directly apply to fine-tune the pretrained Conformer with a multi-scale feature aggregation module, eliminating the need for an additional TDNN- or CNN-based speaker network.
This transfer learning strategy allows the knowledge learned from ASR to be effectively transferred to speaker verification tasks.

\subsection{ASR guided speaker verification}
ASR or phonetic information plays an essential role in speaker verification.
In the statistical i-vector framework, substituting a Gaussian mixture model (GMM) with an ASR-trained DNN to gather sufficient statistics for i-vector extraction results in significant performance improvement \cite{6853887,7404779}.
Alternatively, some researchers utilize a tandem feature that merges spectral and ASR-derived features for GMM modeling \cite{li2016generalized,tian15_interspeech}.

In the realm of deep learning, the integration of ASR and phoneme information into speaker verification is gaining increasing attention. Three main strategies have been investigated for such integration, each with merits and challenges.

The first strategy involves applying frame-level phonetic features from an ASR to a speaker verification model.
In this context, Rahman \textit{et al.} used bottleneck phonetic features from an ASR acoustic model to replace spectral features in speaker network training, indicating the potential of phonetic features to carry speaker-specific information \cite{rahman_employing_2018}.
In similar efforts, researchers have also incorporated phonetic features alongside spectral features for speaker modeling.
Zheng \textit{et al.} used separate network stems to model these two types of features \cite{zheng_phonetically-aware_2020}, while Zhou \textit{et al.} processed these features jointly by concatenating them \cite{zhou_cnn_2019}.
Depending on the modeling stage, phonetic features can be incorporated at the input of the speaker network \cite{zheng_phonetically-aware_2020, zhou_cnn_2019} or before the pooling layer \cite{zhou_cnn_2019, liu_speaker_2018}.
These research indicate that incorporating auxiliary phoneme information benefits speaker modeling.
Besides, Chen \textit{et al.} proposed to model speaker characteristics in phoneme units, termed as phoneme-unit-specific network \cite{chen_phoneme-unit-specific_2021}.
This method can be considered as modeling speaker characteristics using multi-phonetic-head attention, which has the attention weight of phoneme posterior probability.

The second strategy employs a multi-task learning approach, leveraging phoneme recognition as an auxiliary task alongside the primary task of speaker recognition.
Studies have shown that frame-level phoneme modeling enhances speaker verification performance \cite{liu_speaker_2018, tang_collaborative_2017, wang_usage_2019}.

The last strategy involves employing phonetic information as a guided signal to be removed from speaker modeling.
A study by Wang \textit{et al.} suggested that adversarial training to remove phonetic information at the segment level can boost speaker verification performance \cite{wang_usage_2019}.
In contrast, Tawara \textit{et al.} found that removing phonetic information at the frame level is beneficial for extremely short utterances of less than 1.4 seconds \cite{tawara_frame-level_2020-1}.
Hong \textit{et al.} introduced a self-constraint learning and reconstruction strategy that eliminates phonetic information in lower layers, thereby allowing subsequent layers to capture speaker-specific features more efficiently \cite{hong_decomposition_2023}.

In our study, we extend the benefits of the second approach through knowledge distillation from the ASR Conformer to the speaker verification model.
This method aligns with the recognized advantages of employing phoneme recognition as an auxiliary task, thus aiming to improve speaker verification performance.

\subsection{Parameter-efficient transfer learning with adaptors}

The concept of adaptors stems from the idea of fine-tuning large pre-trained models using lightweight neural modules, which can be considered a parameter-efficient transfer learning technique \cite{houlsby_parameter-efficient_2019}.
This approach incorporates trainable lightweight neural modules into a large pre-trained model while keeping the pre-trained parameters frozen during fine-tuning.
This technique has seen successful applications across various domains, including computer vision \cite{rebuffi2017learning}, natural language processing \cite{houlsby_parameter-efficient_2019, he_towards_2022}, and machine translation \cite{bapna_simple_2019}.
 
While adaptors have been successful in different domains, their integration into speech-processing tasks presents multiple applications.
For example, adaptors are applied to self-supervised pre-trained models for speech recognition \cite{thomas_efficient_2022-1}. 
In the context of multilingual ASR, language-specific adaptors have been used to adapt a pre-trained ASR model to various languages \cite{kannan_large-scale_2019, winata_adapt-and-adjust_2021}.
In speech translation, adaptors enable a pre-trained model to specialize in specific language pairs \cite{le_lightweight_2021}.
Additionally, adaptors have been employed to connect an ASR encoder with a multilingual denoising auto-encoder for multilingual speech translation \cite{le_lightweight_2021}.
Other applications of adaptors include speaker verification \cite{peng_parameter-efficient_2023, otake_parameter_2023} and other speech processing tasks \cite{otake_parameter_2023}.

Most existing applications of adaptors focus on self-supervised pre-trained models for specific downstream tasks \cite{houlsby_parameter-efficient_2019, he_towards_2022, peng_parameter-efficient_2023, otake_parameter_2023}.
Moreover, adaptors have been employed to perform domain adaptation for the same task, as seen in multilingual ASR \cite{kannan_large-scale_2019, winata_adapt-and-adjust_2021} and multilingual speech translation \cite{le_lightweight_2021}.
These methods usually incorporate adaptor modules within the network architecture, altering the output of the pre-trained model.

In contrast, our study motivated from the application of the adaptor mechanism.
We apply a similar idea to transfer knowledge across different tasks: from ASR to speaker verification.
We uniquely position an adaptation module on top of the original model, ensuring that the output of the ASR Conformer remains unchanged.
This design enables the simultaneous execution of ASR and speaker verification tasks within a single Conformer model.

\section{methods}
Our research explores three distinct approaches for leveraging an ASR Conformer in speaker verification.
First, we utilize a pre-trained ASR Conformer to initialize the speaker embedding network, which mitigates the risk of overfitting and enhances generalization in the speaker Conformer.
Second, we employ knowledge distillation from the ASR Conformer to the speaker verification model.
Lastly, we introduce an adaptation mechanism that unifies ASR and speaker verification tasks within a single Conformer model. The adaptation efficiently transforms features learned by the ASR to suit speaker verification tasks, all without altering the original ASR Conformer outputs.
This section elaborates on these three methodologies, starting with the architecture of the Conformer encoder.

\subsection{Conformer}
\begin{figure}[t]
  \centering
  \includegraphics[width=0.82\linewidth]{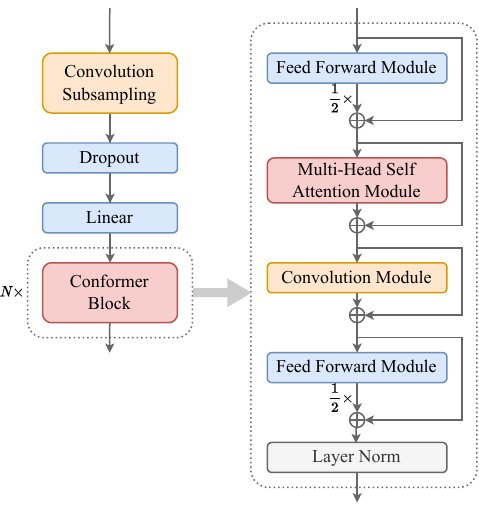}
  \caption{Conformer encoder architecture (left) and a Conformer building block (right) \cite{gulati_conformer_2020}.}
  \label{fig: conformer}
\end{figure}

Developed primarily for ASR tasks, the Conformer encoder is adept at modeling both local and global dependencies within speech signals \cite{gulati_conformer_2020}.
It improves upon the Transformer encoder \cite{vaswani2017attention} by incorporating a CNN to capture local spectral feature information.
The Conformer consists of a convolutional subsampling layer, which reduces the length of input sequences, and a series of Conformer blocks that transform the input signal into higher-level representations.
Fig. \ref{fig: conformer} presents the Conformer encoder structure.

A Conformer block consists of two feed-forward networks (FFNs) flanked by a multi-head self-attention (MHSA) module and a convolution (Conv) module.
In the Conformer, the MHSA employs relative sinusoidal positional encoding \cite{dai2019transformer}, allowing for efficient sequence handling at unseen lengths.
The convolutional module features a point-wise convolution followed by a gated linear unit, succeeded by a one-dimensional depthwise convolution. Batch normalization and Swish activation are subsequently applied.
The feed-forward network contains two linear layers separated by a nonlinear activation, with dropout applied after each linear transformation.
As illustrated in Fig. \ref{fig: conformer}, residual connections are used between the modules, while half-step residual connections are utilized within feed-forward modules, akin to a Macaron-Net \cite{lu2019understanding}.
Layer normalization is applied prior to the output.
Mathematically, for a given input $\mathbf{h}_{i-1}\in \mathbb{R}^{d \times T}$, the output $\mathbf{h}_i\in \mathbb{R}^{d \times T}$ of the $i$-th Conformer block is represented as follows:
\begin{equation}
\begin{aligned}
    \mathbf{h}'_i &= \mathbf{h}_{i-1} + \frac{1}{2}\mathrm{FFN}(\mathbf{h}_{i-1}) \\
    \mathbf{h}''_i &= \mathbf{h}'_i + \mathrm{MHSA}(\mathbf{h}'_i) \\
    \mathbf{h}'''_i &= \mathbf{h}''_i + \mathrm{Conv}(\mathbf{h}''_i) \\
    \mathbf{h}_i &= \mathrm{LayerNorm}(\mathbf{h}'''_i + \frac{1}{2}\mathrm{FFN}(\mathbf{h}'''_i))
\end{aligned}
\end{equation}
where $d$ denotes the dimension of the input and the output sequences, and $T$ represents the length of the time sequence.

\subsection{MFA-Conformer for speaker verification}
Multi-scale feature aggregation (MFA) is a technique that concatenates output feature maps from all frame-level modeling modules in a speaker embedding network before utterance-level pooling.
This approach has been shown to improve performance for TDNN-based networks, suggesting that lower-level features can contribute useful speaker information \cite{desplanques_ecapa-tdnn_2020}.

To apply the Conformer encoder in the speaker verification task, MFA-Conformer proposed to integrate an MFA module into the Conformer encoder \cite{zhang_mfa-conformer_2022}. 
Specifically, this MFA module concatenates the frame-level outputs from all Conformer blocks prior to the pooling layer:
\begin{equation}
\begin{split}
    \mathbf{H}^{\prime} &= \mathrm{Concat}(\mathbf{h}_1, \mathbf{h}_2, \cdots, \mathbf{h}_L)\\
    \mathbf{H} &= \mathrm{LayerNorm}(\mathbf{H}^{\prime})
\end{split}
\end{equation}
where $L$ is the number of Conformer blocks in the Conformer encoder, and $\mathbf{H},\mathbf{H}^{\prime} \in \mathbb{R}^{D \times T}$ with $D = L \times d$.

With this concatenated frame-level feature map $\mathbf{H}$, attentive statistics pooling is applied to produce an utterance-level representation \cite{okabe18_interspeech}.
Finally, the speaker embedding is extracted by applying batch normalization and a fully-connected layer to this utterance-level representation.
During training, an additional fully-connected layer is applied to classify speakers in the training set from speaker embeddings.

\subsection{Transfer learning with the ASR pretrained Conformer}
While deeper Transformers are known to yield superior results as more training data become available \cite{9814838, hsu_hubert_2021-1}, training these models from scratch often requires large datasets \cite{xu-etal-2021-optimizing}.
Further, research indicates that increasing the number of layers in Conformer architectures can result in a performance drop in speaker verification tasks, suggesting potential issues of overfitting \cite{zhang_mfa-conformer_2022}.

To mitigate the risks of overfitting, we employ an ASR pretrained Conformer to initialize the MFA-Conformer-based speaker embedding network.
The pretraining on ASR tasks affords several advantages, such as faster convergence and enhanced generalization capabilities in the speaker verification domain.

In our approach, the parameters of the ASR pretrained Conformer encoder are used to initialize the MFA-Conformer speaker embedding network.
During the early training phases, we keep these encoder parameters frozen and allow only the pooling and subsequent linear layers to be updated for a few epochs.
In later stages, we proceed to fine-tune the parameters across the entire MFA-Conformer architecture to better align it with the specific needs of speaker verification.
By limiting updates to the pooling and linear layers initially, these layers are tailored to adapt the frame-level feature maps derived from the ASR model to the speaker verification objective.
This structured training approach ensures that the pretrained Conformer transitions smoothly to the speaker verification objective without being significantly disrupted by the random initialization of these layers.

\begin{figure}[t]
  \centering
  \includegraphics[width=0.95\linewidth]{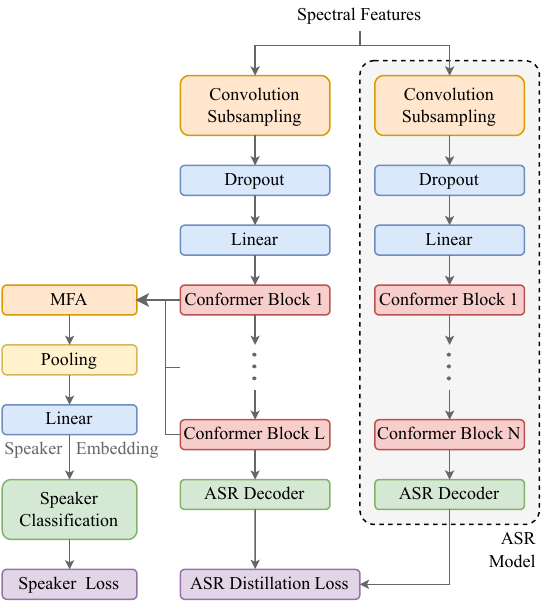}
  \caption{Knowledge distillation from a pretrained ASR Conformer model to a MFA-Conformer-based speaker verification model.}
  \label{fig: distillation}
\end{figure}

\subsection{Knowledge distillation from ASR to speaker verification}
Knowledge distillation involves training a ``student'' model to reproduce the behavior of a more complex ``teacher'' model \cite{hinton_distilling_2015}. 
In our setting, an ASR pretrained Conformer acts as the teacher model, guiding the learning process of the MFA-Conformer-based speaker verification model, which serves as the student.

Given a speaker recognition dataset $\mathcal{D}$, the objective of a speaker verification model is to minimize the difference between its predictions and the ground-truth speaker labels. The loss function $L_{\mathrm{spk}}$ can be expressed as:
\begin{equation}
L_{\mathrm{spk}} = \mathbb{E}_{(\x, y) \sim \mathcal{D}} \left[ \ell_\mathrm{spk}(f(\x), y) \right]
\end{equation}
Here, $f(\cdot)$ is the MFA-Conformer speaker verification model, $f(\x)$ is the Conformer's prediction for the input spectral sequence $\x$, and $y$ is the speaker label. The speaker classification loss $\ell_\mathrm{spk}$ commonly adopts a cross-entropy format or an angular-softmax variant \cite{deng_arcface_2019}.

For distillation, the speaker MFA-Conformer student is trained to align its outputs with the ASR teacher model, as described in the loss $L_{\mathrm{distill}}$:
\begin{equation}
L_{\mathrm{distill}} = \mathbb{E}_{\x \sim \mathcal{D}} \left[ \ell_\mathrm{distill}(f_\mathrm{student}(\x), f_\mathrm{teacher}(\x)) \right]
\end{equation}
In this setting, $f_\mathrm{student}(\cdot)$ refers to the MFA-Conformer coupled with an ASR decoder, while $f_\mathrm{teacher}(\cdot)$ is the ASR model.
\textcolor{black}{In the distillation process, the loss function $L_{\mathrm{distill}}$ is formulated based on the Kullback-Leibler (KL) divergence, which quantify the divergence between the student and teacher frame-level logits outputs.}

The ultimate training objective combines both the speaker classification and the distillation losses:
\begin{equation}\label{eqa: dl}
L = L_{\mathrm{spk}} + \alpha L_{\mathrm{distill}}
\end{equation}
where $\alpha$ is a hyperparameter determining the strength of the distillation effect.
Fig. \ref{fig: distillation} illustrates the knowledge distillation process from ASR to speaker verification.

\textcolor{black}{Our approach harnesses the strengths of both knowledge distillation and multi-task learning, offering advantages for speaker verification.
Firstly, it enables the speaker verification model to utilize robust feature representations from an ASR-pretrained model, enhancing performance without extensive ASR data. This method, diverging from traditional knowledge distillation, incorporates the ASR model's outputs as an auxiliary objective, enriching phonetic feature learning in a multi-task framework.
Secondly, this synergy improves speaker discrimination by leveraging nuanced phonetic information.
Lastly, our method with knowledge distillation offers more architectural flexibility, allowing for optimized designs that can cater to the specific requirements of both ASR and speaker verification tasks.}

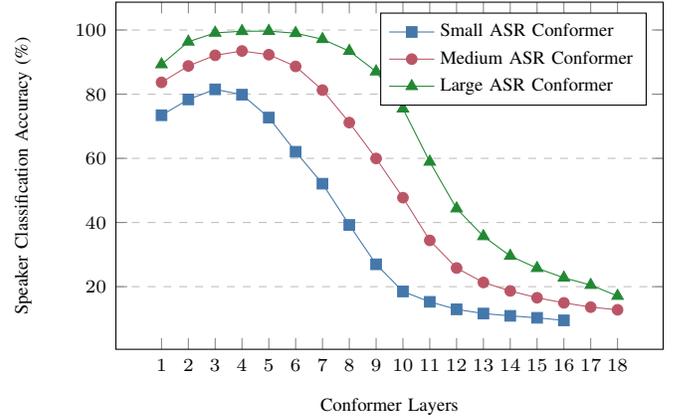
\begin{figure}[t]
  \centering
\begin{tikzpicture}
    \begin{axis}[
        xlabel={Conformer Layers},
        ylabel={Speaker Classification Accuracy (\%)},
        legend pos=north east,
        log ticks with fixed point,
        ymajorgrids=true,
        grid style=dashed,
        width=\linewidth,
        height=0.7\linewidth,
        tick label style={font=\scriptsize},
        legend style={font=\scriptsize, legend cell align={left}},
        label style={font=\scriptsize},
        xtick={1,2,3,4,5,6,7,8,9,10,11,12,13,14,15,16,17,18},
        ytick={20,40,60,80,100},
    ]
    \addplot+[
    	myBlue,
    	mark options={fill=myBlue},
        mark=square*,
    ] coordinates {
        (1,73.4293931034483)(2,78.35934482758626)(3,81.50728275862073)(4,79.84243448275862)(5,72.70206206896547)(6,62.052800000000005)(7,52.090544827586214)(8,39.23623448275861)(9,26.96791034482758)(10,18.50215862068965)(11,15.266717241379311)(12,12.931020689655178)(13,11.61908275862069)(14,10.903855172413795)(15,10.288275862068966)(16,9.492172413793105)
    };
    \addlegendentry{Small ASR Conformer}
    
    \addplot+[
    	myRed,
        mark=*,
        mark options={fill=myRed},
    ] coordinates {
        (1,83.68401724137931)(2,88.79443103448277)(3,92.11478965517233)(4,93.45770344827581)(5,92.30607931034477)(6,88.6529896551724)(7,81.27292068965514)(8,71.14358620689654)(9,59.95691034482763)(10,47.735696551724196)(11,34.41810344827587)(12,25.81356551724139)(13,21.30792413793102)(14,18.665179310344833)(15,16.539568965517244)(16,14.954165517241371)(17,13.612579310344822)(18,12.780175862068964)
    };
    \addlegendentry{Medium ASR Conformer}
    
    \addplot+[
    	myGreen,
        mark=triangle*,
        mark size=2.5pt,
        mark options={fill=myGreen},
    ] coordinates {
        (1,89.26859310344825)(2,96.33482068965522)(3,99.10156551724151)(4,99.60928620689674)(5,99.65372758620703)(6,98.99114137931039)(7,97.12825517241369)(8,93.42808620689658)(9,87.03394482758618)(10,75.48223448275864)(11,58.90494137931032)(12,44.34130344827591)(13,35.69101034482757)(14,29.566296551724122)(15,25.75429655172415)(16,22.76803793103449)(17,20.49298620689656)(18,17.085162068965516)
    };
    \addlegendentry{Large ASR Conformer}
    \end{axis}
\end{tikzpicture}
  \caption{Linear probe accuracy across Conformer layers for speaker identification.}
  \label{fig: spk acc}
\end{figure}

\begin{figure*}[t]
    \centering
    \subfloat[Using frame-level outputs from the $L$-th ASR Conformer layer.\label{fig: adaptor a}]{
    	\includegraphics[width=0.48\linewidth]{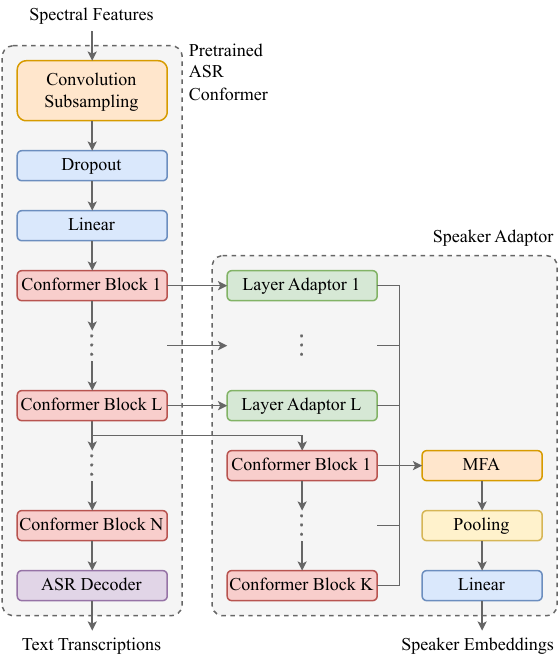}}
    \hfill
    \subfloat[Using concatenated outputs from the first $L$ ASR Conformer layers.\label{fig: adaptor b}]{
    	\includegraphics[width=0.48\linewidth]{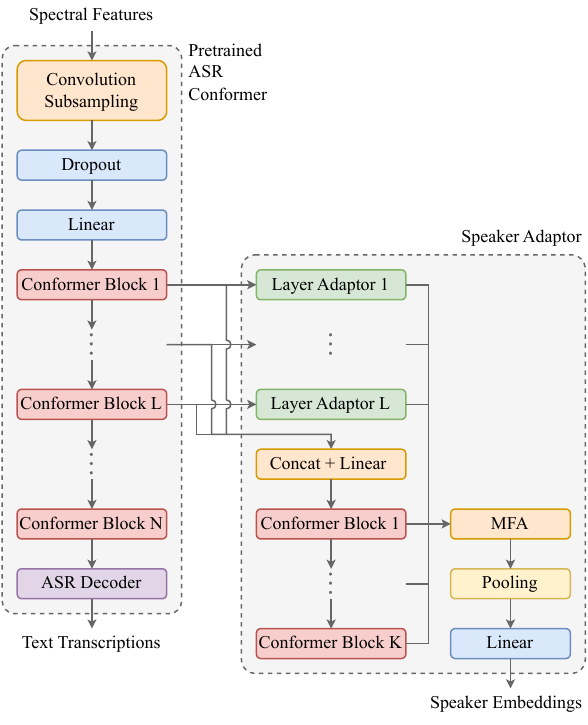}}
    \caption{Design variants of the proposed  \textcolor{black}{speaker adaptation module} to unify ASR and speaker verification in one Conformer model.}
    \label{fig: adaptor}
\end{figure*}

\subsection{\textcolor{black}{Speaker adaptation module}: unifying ASR and speaker verification}\label{sec: adaptor}
To leverage the versatility of Conformer encoders across multiple tasks, this section explores the possibility of crafting a unified model that serves both ASR and speaker verification objectives.

\subsubsection{Inherent speaker-specific information in ASR Conformers}
Conformer encoders, originally tailored for ASR, possess innate adaptability.
This flexibility is attributed to their multi-layered structure, capturing a hierarchical abstraction of speech signals.
Essentially, the lower layers of the ASR Conformer capture diverse attributes of speech, such as speaker characteristics, linguistic patterns, emotional tones, and phonetic variations.
In contrast, the upper layers prioritize phonetic and contextual specifics, driven by the ASR objectives.

To empirically validate this layer-wise specialization, we employed a linear probe to measure the speaker-specific information within different layers of a pretrained ASR Conformer encoder.
A detailed description of the models used for this probing is provided later in section \ref{sec: asr conformer}.
Each Conformer layer's output was first subjected to two linear fully-connected layers, followed by average pooling to derive speaker embeddings. 
These embeddings are further processed by an additional linear layer to perform speaker classification on the VoxCeleb 1 development set \cite{nagrani_voxceleb:_2017}.
The results, illustrated in Fig. \ref{fig: spk acc}, confirm that lower layers inherently possess rich speaker-specific information.
As we progress toward the upper layers, the specificity of the ASR task intensifies, diluting the speaker-specific traits.

\subsubsection{Motivation for a unified Conformer model}
The layer-wise investigation into Conformer encoders revealed an intriguing fact: despite being primarily trained for ASR, even the initial layers possess striking proficiency in speaker recognition.
Remarkably, the fifth layer of a large pretrained ASR Conformer displayed an impressive training accuracy of 99.65\% for speaker recognition, suggesting that ASR-trained features can effectively be used for speaker verification.
This compelling evidence motivates our pursuit of a unified Conformer model that seamlessly transitions between ASR and speaker verification tasks.

\subsubsection{ \textcolor{black}{Speaker adaptation module}}
To bridge the gap between ASR and speaker verification and unify the Conformer encoder, we introduce the speaker adaptation method.
Conceptually, the  \textcolor{black}{speaker adaptation module} is a lightweight trainable module integrated into a large-scale pretrained model \cite{houlsby_parameter-efficient_2019}.
Our design operates on the intermediate representations, leaving the pretrained model's output unchanged.

Fig. \ref{fig: adaptor} visualizes the design of our proposed  \textcolor{black}{speaker adaptation module}.
It consists of three parts: $L$ layer adaptors, $K$ trainable Conformer layers, and a combination of a pooling layer and a subsequent fully connected layer for speaker embedding derivation.

\paragraph{Layer adaptors}
These components work on fine-tuning the outputs from each layer of the pretrained ASR Conformer model, aligning them more closely with the objectives of speaker verification.
Specifically, for a pretrained ASR Conformer, the frame-level output from the $i$-th Conformer layer, denoted as $\h_i \in \mathbb{R}^{d \times T}$, is transformed by the layer adaptor $\A_i$:
\begin{equation}
\h_i' = \mathrm{\A}_i(\h_i)
\end{equation}
Our layer adaptors consist of two linear layers interleaved with layer normalization and an activation function.
Given our observation that deeper layers retain less speaker-centric information, these adaptors are applied only to the first $L$ layers of the pretrained ASR Conformer.

\paragraph{Trainable Conformer layers}
To enhance speaker feature extraction, we incorporate $K$ additional lightweight, trainable Conformer layers within the  \textcolor{black}{speaker adaptation module}.
Inputs to these layers come from one of the two following distinct options:
\begin{itemize}
	\item Frame-level outputs from the $L$-th Conformer layer of the ASR model, as illustrated in Fig. \ref{fig: adaptor a}.
	\item Concatenated outputs from the first $L$ layers of the pretrained ASR Conformer encoder, with a linear layer to reduce the feature dimension, as illustrated in Fig. \ref{fig: adaptor b}.
\end{itemize}
To maintain the efficiency of the  \textcolor{black}{speaker adaptation module}, these trainable Conformer layers are designed to be lightweight, with reduced hyper-parameters of dimensions and hidden units.

\paragraph{Speaker embedding extraction}
After the transformations brought by the layer adaptors and the trainable Conformer layers, the frame-level features are fed into the MFA module:
\begin{equation}
\begin{split}
    \mathbf{H}^{\prime} &= \mathrm{Concat}[\mathrm{\A}_1(\h_1), \cdots, \mathrm{\A}_L(\h_L), \tilde{\h}_1, \cdots,\tilde{\h}_K]\\
    \mathbf{H} &= \mathrm{LayerNorm}(\mathbf{H}^{\prime})
\end{split}
\end{equation}
Here, $\tilde{\h}_k$ denotes the output from the $k$-th trainable Conformer layer. $K$ represents the number of these layers.
By design, $K$ can be zero, indicating the absence of any new trainable Conformer layers.
With these concatenated frame-level representations derived from the pretrained ASR Conformer encoder, a standard speaker verification procedure with an utterance-level pooling layer and a subsequent linear layer is used for speaker embedding extraction.

During the training phase, the pretrained ASR Conformer is kept frozen. 
Only  \textcolor{black}{speaker adaptation module} components, including layer adaptors, lightweight Conformer layers, pooling, and the following linear layers, are trained under the speaker verification objective.

\section{Experimental Setups}

\subsection{Dataset}
The experiments are conducted on VoxCeleb \cite{nagrani_voxceleb:_2017, chung_voxceleb2:_2018}.
For model training, we opted to employ the development set from VoxCeleb 2. This training dataset encompasses 1,092,009 audio recordings from a diverse set of 5,994 distinct speakers.

For the evaluation phase, we use both the development and test sets from VoxCeleb 1. We present the speaker verification performances based on three predefined trial lists as described in \cite{chung_voxceleb2:_2018}:
\begin{itemize}
\item VoxCeleb 1-O: This represents the original trial list associated with VoxCeleb 1, encompassing 37,720 trials derived from 40 speakers.
\item VoxCeleb 1-E: An expanded trial list that comprises 581,480 trials sourced from 1,251 speakers.
\item VoxCeleb 1-H: A more challenging trial list with 552,536 trials from 1,190 speakers. All test pairings within this list share the same linguistic background and gender.
\end{itemize}

\begin{table}[t]
  \caption{Three ASR Conformer encoders of different sizes}
  \label{tab:netarc}
  \centering
  \begin{tabular}[c]{@{\ }l@{\ \ \ }c@{\ \ \ }c@{\ \ \ }c@{\ \ \ }c@{\ \ \ }c@{\ }}
    \toprule 
    Model & layers & dim & heads & hidden units & parameters\\
    \midrule
    Small\tablefootnote{\url{https://catalog.ngc.nvidia.com/orgs/nvidia/teams/nemo/models/stt_en_conformer_ctc_small}} & 16 & 176 & 4 & 704 & 15.88M\\
    Medium\tablefootnote{\url{https://catalog.ngc.nvidia.com/orgs/nvidia/teams/nemo/models/stt_en_conformer_ctc_medium}} & 18 & 256 & 4 & 1024 & 35.26M\\
    Large\tablefootnote{\url{https://catalog.ngc.nvidia.com/orgs/nvidia/teams/nemo/models/stt_en_conformer_ctc_large}} & 18 & 512 & 8 & 2048 & 130.94M\\
    \bottomrule
  \end{tabular}
\end{table}

\subsection{Data Augmentation}
To enhance the robustness and versatility of our model, we integrated various data augmentation methodologies.
First, we apply speed perturbation to the audio samples by accelerating or decelerating the content by factors of 1.1 and 0.9, respectively \cite{yamamoto19_interspeech, wang2020dku}. 
As a result, this approach produced two supplementary replicas of each original audio, expanding the entire training dataset to include 17,982 distinct speakers and 3,276,027 unique utterances.

For the enlarged training dataset, two primary strategies were utilized:
\begin{itemize}
\item Additive noise augmentation: The MUSAN dataset \cite{musan} served as our noise source, enabling us to add ambient noise, musical sounds, and babble noise onto our audio files.
The babble noise was generated by merging between three to eight separate speech files in the MUSAN dataset.
The signal-to-noise ratios (SNR) range from 0 to 20 dB.
\item Convolutional reverberation noise augmentation: We employed the collection of 40,000 simulated room impulse responses (RIR) from the study in \cite{ko2017study}. Only simulated RIRs originating from small to medium-sized rooms are used.
\end{itemize}
To maintain variability during training epochs, we integrated on-the-fly data augmentation, applying the aforementioned noise augmentations with a likelihood of 0.6 for each training speech.

\subsection{Pretrained ASR Conformer}
\label{sec: asr conformer}
We utilize pretrained ASR models from the NEMO toolkit \cite{kuchaiev2019nemo}.
The choice of the Conformer model from NEMO was driven by its performance and generalization capabilities, as demonstrated in various benchmarks.
This ASR Conformer adopts the same encoder architecture as illustrated in \cite{gulati_conformer_2020} but uses a linear decoder and the connectionist temporal classification (CTC) for decoding. 

In our experiments, we use three sizes of the NEMO ASR Conformer: small, medium, and large.
Despite variations in size, each of these models shares a convolution subsampling rate of $\frac{1}{4}$, along with a consistent kernel size of 31 for their convolution modules.
Table \ref{tab:netarc} shows the differences in Conformer layer numbers, encoder dimensions, attention heads, and linear hidden units across the three Conformer encoders.

According to the NEMO toolkit documentation, each Conformer-CTC model is trained on English corpora collated from 10 distinct datasets.\footnote{These datasets include Librispeech, Fisher Corpus, Switchboard-1, WSJ-0 and WSJ-1, National Speech Corpus (Part 1, Part 6), VCTK, VoxPopuli (EN), Europarl-ASR (EN), Multilingual Librispeech (MLS EN 2,000 hours subset), and Mozilla Common Voice (v7.0).}
In total, this collection spans approximately 10,000 hours of speech data.\footnote{This estimate is derived from the training data descriptions provided at the mentioned link in the previous footnote.}

\begin{table*}[t]
  \caption{Speaker verification performance of ASR pretrained MFA-Conformer on VoxCebleb 1.}
  \label{tab:result1}
  \centering
  \begin{tabular}[c]{lcc cccccc}
    \toprule
    \multirow{2}*{\textbf{Model}} & \multirow{2}*{\textbf{Size}} & \multirow{2}*{\textbf{Pretrained}} & \multicolumn{2}{c}{\textbf{VoxCeleb 1-O}} & \multicolumn{2}{c}{\textbf{VoxCeleb 1-E}} & \multicolumn{2}{c}{\textbf{VoxCeleb 1-H}} \\
    \cmidrule(lr){4-5} \cmidrule(lr){6-7} \cmidrule(lr){8-9}
    & & & EER[\%] & minDCF & EER[\%] & minDCF & EER[\%] & minDCF \\
    \midrule
    ECAPA-TDNN \cite{9414600} & 46.6M & $\times$ & 0.68 & 0.0753 & 0.91 & 0.1006 & 1.72 & 0.1695 \\
    HuBERT Large \cite{chen_large-scale_2022} & 316.61M+ & $\surd$ & 0.72 & - & 0.70 & - & 1.32 & - \\
    Wav2Vec2.0 Large (XLSR) \cite{chen_large-scale_2022} & 317.38M+ & $\surd$ & 0.73 & - & 0.68 & - & 1.23 & - \\
    UniSpeech-SAT Large \cite{chen_large-scale_2022} & 316.61M+ & $\surd$ & 0.63 & - & 0.63 & - & 1.29 & - \\
    WavLM Large + QMF \cite{9814838} & 316.62M+ & $\surd$ & 0.38 & - & 0.48 & - & 0.99 & - \\ 
    \midrule
    NEMO Small & 15.88M & $\times$ & \textbf{0.88} & \textbf{0.1367} & \textbf{1.08} & \textbf{0.1342} & \textbf{2.20} & \textbf{0.2245} \\
    NEMO Medium & 35.26M & $\times$ & 0.94 & 0.1200 & 1.26 & 0.1487 & 2.41 & 0.2398 \\
    NEMO Large & 130.94M & $\times$ & 0.96 & 0.1375 & 1.22 & 0.1391 & 2.35 & 0.2278 \\
    \midrule
    NEMO Large first 4 layers & 35.02M & $\times$ & 0.86 & 0.1051 & 1.03 & 0.1188 & 1.97 & 0.1920 \\
    NEMO Large first 6 layers & 48.72M & $\times$ & 0.80 & 0.1101 & 1.04 & 0.1202 & 2.04 & 0.2012 \\
    NEMO Large first 8 layers & 62.42M & $\times$ & 0.81 & 0.1121 & 1.00 & 0.1183 & 1.93 & 0.1904 \\
    \midrule
    NEMO Small & 15.88M & $\surd$ & 0.74 & 0.1101 & 0.90 & 0.1054 & 1.90 & 0.1893 \\
    NEMO Medium & 35.26M & $\surd$ & 0.61 & 0.0946 & 0.78 & 0.0891 & 1.67 & 0.1649 \\
    NEMO Large & 130.94M & $\surd$ & \textbf{0.48} & \textbf{0.0673} & \textbf{0.71} & \textbf{0.0785} & \textbf{1.54} & \textbf{0.1538} \\
    \ \ \ \ + QMF & & & \textbf{0.43} & \textbf{0.0623} & \textbf{0.66} & \textbf{0.0709} & \textbf{1.35} & \textbf{0.1350} \\

    \midrule
    NEMO Large first 4 layers & 35.02M & $\surd$ & 0.77 & 0.1065 & 1.04 & 0.1159 & 1.95 & 0.1862 \\
    NEMO Large first 6 layers & 48.72M & $\surd$ & 0.58 & 0.0618 & 0.84 & 0.0937 & 1.62 & 0.1571 \\
    NEMO Large first 8 layers & 62.42M & $\surd$ & 0.64 & 0.0982 & 0.86 & 0.0944 & 1.77 & 0.1732 \\
    \bottomrule
  \end{tabular}
\end{table*}

\subsection{Implementation details}

Speech utterances are cropped to 2 seconds for training the speaker embedding network. 
We use a logarithmic Mel-spectrogram with 80 frequency bins as the acoustic feature, computed over Hamming windows of 20ms with a 10ms shift.

During training, the Additive angular margin (AAM) loss \cite{deng_arcface_2019} is employed with a re-scaling factor of 32 and an angular margin of 0.2 to learn discriminative representations.
The speaker embedding dimension is set to 256.
We utilize the AdamW optimizer, beginning with a learning rate of 0.001.
Additionally, we implement a cosine annealing learning rate scheduler, incorporating a warm-up phase spanning one training epoch.
Our chosen batch size is 512, with a weight decay of $10^{-7}$.

After convergence, we employ large margin fine-tuning (LMFT) \cite{9414600}.
Speech segments are expanded to 6 seconds, and the angular margin in the AAM loss is increased to 0.5.
We turn off speed perturbation data augmentation, reverting the training data to its original set.

\subsection{Evaluation}
To generate speaker verification scores, we apply the adapted score normalization \cite{sn_analysis_2017} after cosine similarity on two given speaker embeddings.
In adapted score normalization, we utilize an imposter cohort randomly chosen from 30,000 training utterances, with an adapted cohort size of 700.

Although our standard procedure involves only this score normalization, we further calibrate the verification scores using the Quality Measure Function (QMF) \cite{6584746, 9414600} for specific systems as per their requirements.
The calibration model is trained on 30,000 trials generated from the VoxCeleb 2 development set. This model incorporates several quality metrics including the duration and SNR of the enrollment and testing utterances, the magnitudes of the embeddings, and the verification score itself.

We evaluate speaker verification performance using two metrics: (1) Equal Error Rate (EER): This denotes the error rate at the point where the false acceptance rate equals the false rejection rate. (2) Minimum Detection Cost (minDCF): This represents the minimal value of a detection cost function. The function is a weighted sum of false-reject and false-alarm error rates for a given decision threshold \cite{nist_nist_2016}. The parameters for this function are set as follows: $C_{\mathrm{Miss}}=1$, $C_{\mathrm{FA}}=1$, and $P_{\mathrm{Target}}=0.01$.

\section{Experimental Results}

\subsection{Transfer learning with the ASR pretrained Conformer}\label{exp1}
In this subsection, we present speaker verification results using our first proposed method.
Specifically, we explore the efficacy of initializing the MFA-Conformer speaker verification model with a pretrained ASR Conformer.
The performance of various MFA-Conformer speaker embedding networks, both with and without ASR pretraining, are detailed in Table \ref{tab:result1}.

\subsubsection{MFA-Conformer's performance without ASR pretraining}
We first analyze the performance of the MFA-Conformer model without integrating ASR pretraining.
The results indicate that increasing the trainable parameters does not yield improved speaker verification performance.
Specifically, upon increasing model parameters by a factor of eight (from 15.88 million to 130.94 million), the EERs observe a decrease ranging from 7\% to 13\% across the three testing trials.
This suggests that MFA-Conformers tend to overfit, especially in scenarios with limited data availability.

\begin{table*}[t]
     \caption{Comparsion of ASR pretraining method and SSL as front-end module method. Performance are reported on EER (\%).}
    \label{tab:comp_asr_ssl_}
    \centering
   \begin{tabular}{@{\ \ }l@{\ \ }c@{\ \ }c@{\ \ }c@{\ \ }c@{\ \ }c@{\ \ }c@{\ \ }c@{\ \ }c@{\ \ }c@{\ \ }}
    \toprule
        \multicolumn{4}{c}{Pretrained Model} & \multirow{2}*{Speaker Model} & \multirow{2}*{LMFT} & \multirow{2}*{QMF} & \multirow{2}*{Vox1-O} & \multirow{2}*{Vox1-E} & \multirow{2}*{Vox1-H} \\
    \cmidrule(lr){1-4}
        Model & Size & Training Data & Usage \\
    \midrule
         HuBERT Base \cite{9814838} & 94.7M & 960 hr & front-end module & ECAPA-TDNN & $\times$ & $\times$ & 0.989 & 1.068 & 2.216 \\
         HuBERT Large \cite{9814838} & 316.6M & 60k hr & front-end module & ECAPA-TDNN & $\times$ & $\times$ & 0.808 & 0.822 & 1.678 \\
         HuBERT Large \cite{9814838} & 316.6M & 60k hr & front-end module & ECAPA-TDNN & $\surd$ & $\surd$ & 0.585 & 0.654 & 1.342 \\
\midrule
         WavLM Base+ \cite{9814838} & 94.7M & 94k hr & front-end module & ECAPA-TDNN & $\times$ & $\times$ & 0.84 & 0.928 & 1.758 \\
         WavLM Large \cite{9814838} & 316.6M & 94k hr & front-end module & ECAPA-TDNN & $\times$ & $\times$ & 0.617 & 0.662 & 1.318 \\
         WavLM Large \cite{9814838} & 316.6M & 94k hr & front-end module & ECAPA-TDNN & $\surd$ & $\surd$ & 0.383 & 0.480 & 0.986 \\
\midrule
        Conformer Medium & 35.3M & 10k hr & parameter initialization & pretrained Conformer & $\times$ & $\times$ & 0.78 & 0.97 & 2.04 \\
        Conformer Medium & 35.3M & 10k hr & parameter initialization & pretrained Conformer & $\surd$ & $\times$ & 0.61 & 0.78 & 1.67 \\
        Conformer Medium & 35.3M & 10k hr & parameter initialization & pretrained Conformer & $\surd$ & $\surd$ & 0.52 & 0.72 & 1.48 \\
\midrule
        Conformer Large & 130.9M & 10k hr & parameter initialization & pretrained Conformer & $\times$ & $\times$ & 0.74 & 0.91 & 1.91 \\
        Conformer Large & 130.9M & 10k hr & parameter initialization & pretrained Conformer & $\surd$ & $\times$ & 0.48 & 0.71 & 1.54 \\
        Conformer Large & 130.9M & 10k hr & parameter initialization & pretrained Conformer & $\surd$ & $\surd$ & 0.43 & 0.66 & 1.35\\
    \bottomrule
    \end{tabular}
\end{table*}

\subsubsection{MFA-Conformer's performance with ASR pretraining}
Integrating ASR pretraining into the MFA-Conformer model leads to significant improvements across all evaluated model sizes.
For example, the small MFA-Conformer with ASR pretraining recorded a relative reduction in EER of 15.9\% on the VoxCeleb 1-O trails compared to its non-pretrained counterpart.
This relative reduction was even more significant for larger models, with the large MFA-Conformer recording a 50\% reduction on the same trail.
\textcolor{black}{These results confirm the benefits of leveraging ASR pretraining with 10k hours of speech data for speaker verification models, particularly for larger Conformer models, where the risk of overfitting is higher.}

\begin{table*}[h]
  \caption{Speaker verification performance of MFA-Conformer with ASR distillation on VoxCebleb 1.}
  \label{tab:result2}
  \centering
  \begin{tabular}[c]{lccc lccccccc}
    \toprule
    \multirow{2}*{\textbf{Model}} & \multirow{2}*{\makebox[1cm]{\parbox{1cm}{\centering \textbf{Sampling\\Rate}}}} & \multirow{2}*{\textbf{Size}} & \multirow{2}*{\textbf{MACs}\footnotemark} & \multirow{2}*{\makebox[1cm]{\parbox{1cm}{\textbf{Training\\Method}}}} & \multicolumn{2}{c}{\textbf{VoxCeleb 1-O}} & \multicolumn{2}{c}{\textbf{VoxCeleb 1-E}} & \multicolumn{2}{c}{\textbf{VoxCeleb 1-H}} \\
    \cmidrule(lr){6-7} \cmidrule(lr){8-9} \cmidrule(lr){10-11}
    & & & & & EER[\%] & minDCF & EER[\%] & minDCF & EER[\%] & minDCF \\
    \midrule
    \multirow{3}*{\makebox[1.55cm]{\parbox{1.55cm}{NEMO\\Half Small}}} & \multirow{3}*{$\frac{1}{2}$} & \multirow{3}*{8.73M} & \multirow{3}*{405.18M} & Baseline & 0.62 & 0.0792 & 0.84 & 0.0907 & 1.67 & 0.1676 \\
     & & & & ASR Distillation & 0.65 & 0.0725 & 0.79 & 0.0881 & 1.50 & 0.1477 \\
& & & & \ \ \ \ + QMF & \textcolor{gray}{0.56} & \textcolor{gray}{0.0572} & \textcolor{gray}{0.74} & \textcolor{gray}{0.0775} & \textcolor{gray}{1.36} & \textcolor{gray}{0.1333} \\
    \midrule
    \multirow{5}*{\makebox[1.55cm]{\parbox{1.55cm}{NEMO\\Small}}} & \multirow{5}*{$\frac{1}{4}$} & \multirow{5}*{15.88M} & \multirow{5}*{1.12G} & Baseline & 0.88 & 0.1367 & 1.08 & 0.1342 & 2.20 & 0.2245 \\
     & & & & ASR Pretrained & 0.74 & 0.1101 & 0.90 & 0.1054 & 1.90 & 0.1893 \\
& & & & \ \ \ \ + QMF & \textcolor{gray}{0.61} & \textcolor{gray}{0.0937} & \textcolor{gray}{0.83} & \textcolor{gray}{0.0954} & \textcolor{gray}{1.69} & \textcolor{gray}{0.1687} \\
     & & & & ASR Distillation & 0.54 & 0.0625 & 0.74 & 0.0782 & 1.54 & 0.1568 \\
& & & & \ \ \ \ + QMF & \textcolor{gray}{0.43} & \textcolor{gray}{0.0575} & \textcolor{gray}{0.67} & \textcolor{gray}{0.0705} & \textcolor{gray}{1.37} & \textcolor{gray}{0.1429} \\
    \midrule
    \multirow{3}*{\makebox[1.55cm]{\parbox{1.55cm}{NEMO\\Half Medium}}} & \multirow{3}*{$\frac{1}{2}$}  & \multirow{3}*{19.30M} & \multirow{3}*{803.04M} & Baseline & 0.64 & 0.0855 & 0.89 & 0.1020 & 1.74 & 0.1750 \\
    & & & & ASR Distillation & \textbf{0.43} & \textbf{0.0485} & \textbf{0.69} & \textbf{0.0727} & \textbf{1.37} & \textbf{0.1364} \\
& & & & \ \ \ + QMF & \textcolor{gray}{0.38} & \textcolor{gray}{0.0388} & \textcolor{gray}{0.66} & \textcolor{gray}{0.0668} & \textcolor{gray}{1.24} & \textcolor{gray}{0.1221} \\
    \midrule
    \multirow{5}*{\makebox[1.55cm]{\parbox{1.55cm}{NEMO\\Medium}}} & \multirow{5}*{$\frac{1}{4}$}  & \multirow{5}*{35.26M} & \multirow{5}*{2.31G} & Baseline & 0.94 & 0.1200 & 1.26 & 0.1487 & 2.41 & 0.2398 \\
    & & & & ASR Pretrained & 0.61 & 0.0946 & 0.78 & 0.0891 & 1.67 & 0.1649 \\
& & & & \ \ \ \ + QMF & \textcolor{gray}{0.52} & \textcolor{gray}{0.0875} & \textcolor{gray}{0.72} & \textcolor{gray}{0.0783} & \textcolor{gray}{1.48} & \textcolor{gray}{0.1538} \\
    & & & & ASR Distillation & \textbf{0.52} & \textbf{0.0689} & \textbf{0.72} & \textbf{0.0791} & \textbf{1.49} & \textbf{0.1429} \\
& & & & \ \ \ \ + QMF & \textcolor{gray}{0.48} & \textcolor{gray}{0.0589} & \textcolor{gray}{0.67} & \textcolor{gray}{0.0711} & \textcolor{gray}{1.34} & \textcolor{gray}{0.1364} \\
    \midrule
    \multirow{3}*{\makebox[1.55cm]{\parbox{1.55cm}{NEMO\\Half Large}}} & \multirow{3}*{$\frac{1}{2}$} & \multirow{3}*{72.16M} & \multirow{3}*{2.52G} & Baseline & 0.87 & 0.0799 & 1.04 & 0.1145 & 1.93 & 0.1838 \\
    & & & & ASR Distillation & 0.52 & 0.0564 & 0.75 & 0.0808 & 1.55 & 0.1516 \\
& & & & \ \ \ \ + QMF & \textcolor{gray}{0.48} & \textcolor{gray}{0.0619} & \textcolor{gray}{0.72} & \textcolor{gray}{0.0735} & \textcolor{gray}{1.42} & \textcolor{gray}{0.1439} \\
    \midrule
    \multirow{5}*{\makebox[1.55cm]{\parbox{1.55cm}{NEMO\\Large}}} & \multirow{5}*{$\frac{1}{4}$}  & \multirow{5}*{130.94M} & \multirow{5}*{8.53G} & Baseline & 0.96 & 0.1375 & 1.22 & 0.1391 & 2.35 & 0.2278 \\
    & & & & ASR Pretrained & \textbf{0.48} & \textbf{0.0673} & \textbf{0.71} & \textbf{0.0785} & \textbf{1.54} & \textbf{0.1538} \\
& & & & \ \ \ \ + QMF & \textcolor{gray}{0.43} & \textcolor{gray}{0.0623} & \textcolor{gray}{0.66} & \textcolor{gray}{0.0709} & \textcolor{gray}{1.35} & \textcolor{gray}{0.1350} \\
    & & & & ASR Distillation & 0.53 & 0.0589 & 0.79 & 0.0852 & 1.64 & 0.1611 \\
& & & & \ \ \ \ + QMF & \textcolor{gray}{0.45} & \textcolor{gray}{0.0562} & \textcolor{gray}{0.75} & \textcolor{gray}{0.0802} & \textcolor{gray}{1.49} & \textcolor{gray}{0.1475} \\
   \bottomrule
  \end{tabular}
\end{table*}

\subsubsection{Benchmarking against large self-supervised speech models}
\textcolor{black}{
Large self-supervised speech models for speaker verification are used as feature extractors to replace the handcrafted feature with an additional speaker embedding model append.
}
Compared to larger self-supervised pretrained models with more than 300 million parameters (HuBERT Large, Wav2Vec2.0 Large, UniSpeech-SAT Large), the ASR pretrained MFA-Conformers achieve comparable or even better verification performance on VoxCeleb 1-O trials. 
For instance, while the UniSpeech-SAT large model (with 316.62 million parameters) achieved an EER of 0.63\% on VoxCeleb 1-O trials, the large ASR pretrained MFA-Conformer (with 130.94 million parameters) recorded an EER of 0.48\%.
Such results emphasize the efficiency of ASR pretraining on the speaker MFA-Conformer model.

However, MFA-Conformers do not outperform large self-supervised models on the VoxCeleb 1-E and VoxCeleb 1-H trials.
A plausible reason is the difference in the volume of training data used for pretraining.
While self-supervised models utilized speech data ranging from 56,000 to 188,000 hours, the training data of the ASR Conformer used in this study are limited to approximately 10,000 hours.
Nevertheless, our proposed ASR pretraining method offers flexibility.
Integrating an MFA module and a pooling layer can readily transform an ASR pretrained Conformer into a speaker verification task.
This eliminates the need for supplementary TDNN- or CNN-based speaker networks, which are commonly employed in large self-supervised models.

\textcolor{black}{
To facilitate a direct comparison between the ASR pretrained method and the large self-supervised speech model method, table \ref{tab:comp_asr_ssl_} highlights various configurations, including different model sizes, training data, usage types, and additional fine-tuning techniques like Large Margin Fine-Tuning (LMFT) and Quality Measure Function (QMF).  From the table, the medium Conformer model with ASR pretraining demonstrates comparable performance to the WavLM Base+ and HuBERT Base models. Similarly, the large ASR pretrained Conformer model exhibits performance on par with the HuBERT Large and WavLM Large models using a smaller size of model and training data, making it a competitive option in the realm of speech model methods.
}

\subsubsection{Exploring the potential of extracting lower layers}
We also extend our experiments by using subsets of the larger Conformer model, specifically extracting the initial 4, 6, and 8 layers, to initiate MFA-Conformer training.
These truncated models perform better than the full version when ASR pretraining was not applied, which reaffirms the earlier observation regarding the overfitting tendency of Conformers with increased parameters.
When ASR pretraining is applied, these truncated models outperform their counterparts without ASR pretraining, emphasizing the benefits of ASR pretraining.
The experiments of the truncated Conformers present a way to balance model size and speaker verification performance.

\subsection{Knowledge distillation from ASR to speaker verification}\label{exp2}

This section presents the results of our second proposal, which explores the application of knowledge distillation from ASR to speaker verification.
For these experiments, we used the NEMO Large ASR-CTC model in Table \ref{tab:netarc}, as the teacher model in the knowledge distillation process.
We set the hyperparameter $\alpha$ in equation \ref{eqa: dl} to 1. 
The speaker verification performance of various MFA-Conformer models, considering different training methodologies and model sizes, are shown in Table \ref{tab:result2}.

\subsubsection{Influence of ASR knowledge distillation}
The primary objective of our experiments is to determine the effectiveness of ASR distillation in enhancing the performance of MFA-Conformer models.
The application of ASR distillation consistently shows promising improvements across various model scales and sampling rates.
For instance, considering the NEMO Small model, the ASR distillation technique (EER of 0.54\%) reduces the EER by 38.6\% on the VoxCeleb 1-O trials compared to the vanilla version (EER of 0.88\%).
The NEMO Medium model with ASR distillation outperforms its vanilla counterpart by approximately 44.7\% relatively in EER on the same trials.

Our results also enable a direct comparison between the ASR distillation and ASR pretraining techniques.
Notably, in most cases, models trained with ASR distillation outperform or come close to their ASR pretrained counterparts. 
For instance, the EER in the VoxCeleb 1-O trial for the NEMO Medium model decreases by 14.8\% with ASR distillation compared to ASR pretraining.
The improvements from ASR distillation primarily come from two factors.
First, the student model benefits from the robustness of the larger ASR teacher model trained on extensive ASR datasets, exposing the student model to a wide range of speech patterns and accents.
Second, the auxiliary task of ASR at frame-level modeling enhances the student model's ability to capture fine-grained, speaker-specific features, which is critical for speaker verification.

However, the NEMO Large model with ASR distillation does not consistantly outperform the ASR pretraining method.
This might be due to the shared model architecture between the student and teacher models, as the ASR-pretrained NEMO Large model was used as the teacher.
This outcome suggests no one-size-fits-all answer, and the best approach could depend on the specific model architecture or data constraints.

\subsubsection{Reduced Conformer layers with increased \textcolor{black}{convolution subsampling rate}}
To explore the impact of model size and sampling rate, we reduced the number of Conformer layers by half and increased the \textcolor{black}{convolution subsampling rate} from $\frac{1}{4}$ to $\frac{1}{2}$ for the three Conformer models.
The ASR teacher model remained the same as in previous experiments. 
To match the \textcolor{black}{convolution subsampling rate} between the teacher and student models for the KL divergence loss at frame level, we added a convolutional layer to the student model with a kernel size of 3, padding of 1, and stride of 2, increasing the student model's \textcolor{black}{convolution subsampling rate} from $\frac{1}{4}$ to $\frac{1}{2}$.

Our results show that MFA-Conformer models with a $\frac{1}{2}$ \textcolor{black}{convolution subsampling rate}, even with nearly half the number of parameters, achieve comparable or better verification performance with ASR distillation compared to those with a $\frac{1}{4}$ \textcolor{black}{convolution subsampling rate}.
For example, the NEMO Half Medium model with ASR distillation achieved EERs of 0.43\%, 0.69\%, and 1.37\%, while the NEMO Medium model's EERs were 0.52\%, 0.72\%, and 1.49\% for VoxCeleb 1-O, VoxCeleb 1-E, and VoxCeleb 1-H trials, respectively.

The integration of ASR distillation into the MFA-Conformer model training presents a promising direction in speaker verification. 
Our results demonstrate consistent improvements across different model scales, indicating the robustness and versatility of this method. 
Moreover, the potential to achieve similar or even better results than ASR pretraining further highlights the efficacy of ASR distillation.

\begin{table}[t]
\caption{The network configuration of the  \textcolor{black}{speaker adaptation module}.}
\label{tab: adapter arch}
\centering
\begin{tabular}[c]{l l|l}
\toprule
\textbf{Layer} & \multicolumn{2}{l}{\textbf{Structure}} \\
\midrule
\tabincell{l}{Layer\\Adaptor} & \multicolumn{2}{l}{$\left[ \begin{array}{l} \text{Linear}(D, 128) \\ \text{LayerNorm}(128) \\ \text{ReLU}() \\ \text{Linear}(128, 128) \end{array} \right]\times L$} \\
\midrule
\multirow{3}*{\tabincell{l}{Trainable\\Conformer}} & V2: Linear$(D, 176)$ & 
V3: \tabincell{l}{Concatenation\\Linear$(D\times L, 176)$} \\
\cmidrule{2-3}
& \multicolumn{2}{l}{Conformer$(\text{dim}=176, \text{head}=4, \text{hidden}=704)\times K$} \\
\midrule
\multirow{2}*{MFA} & \multicolumn{2}{l}{Concatenation} \\
& \multicolumn{2}{l}{LayerNorm$(128\times L + 176\times K)$} \\
\midrule
Pooling & \multicolumn{2}{l}{Attentive statistics pooling} \\
\midrule
Linear & \multicolumn{2}{l}{Linear$((128\times L + 176\times K)\times2, 256)$} \\
\bottomrule
\end{tabular}
\end{table}

\footnotetext{MACs (Multiply-Accumulate Operations) are calculated based on a 5-second speech input.}

\begin{table*}[t]
  \caption{EER[\%] and minDCF of the speaker adaptation method on VoxCebleb 1.}
  \label{tab:result6}
  \centering
  \begin{tabular}[c]{lcccccccccc}
    \toprule
    \multirow{2}*{\textbf{Model}} & \multicolumn{2}{@{}c@{}}{\textbf{Size}} & \multicolumn{2}{@{}c@{}}{\textbf{MACs}} & \multicolumn{2}{@{}c@{}}{\textbf{Vox1-O}} & \multicolumn{2}{@{}c@{}}{\textbf{Vox1-E}} & \multicolumn{2}{@{}c@{}}{\textbf{\ Vox1-H}} \\
    \cmidrule(lr){2-3} \cmidrule(lr){4-5} \cmidrule(lr){6-7} \cmidrule(lr){8-9} \cmidrule(lr){10-11}
    & ASR & SpkAdap & ASR & SpkAdap & EER & minDCF & EER & minDCF & EER & minDCF \\
    \midrule
    Small V3 $L$8 $K$2 & \multirow{2}*{6.94M} & \multirow{2}*{3.49M} & \multirow{2}*{826.39M} & \multirow{2}*{116.51M} & 0.83 & 0.1223 & 0.99 & 0.1058 & 1.87 & 0.1798 \\
	\ \ \ \ + QMF & & & & & 0.69 & 0.1011 & 0.89 & 0.0930 & 1.66 & 0.1663 \\
	\midrule
    Medium V3 $L$10 $K$2 & \multirow{2}*{17.79M} & \multirow{2}*{4.14M} & \multirow{2}*{1.79G} & \multirow{2}*{116.71M} & 0.67 & 0.0873 & 0.88 & 0.0978 & 1.66 & 0.1609 \\
	\ \ \ \ + QMF & & & & & 0.55 & 0.0807 & 0.80 & 0.0844 & 1.48 & 0.1494 \\
	\midrule
    Large V3 $L$10 $K$2 & \multirow{2}*{70.85M} & \multirow{2}*{4.92M} & \multirow{2}*{7.07G} & \multirow{2}*{117.33M} & 0.57 & 0.0631 & 0.77 & 0.0805 & 1.52 & 0.1484 \\
	\ \ \ \ + QMF & & & & & 0.45 & 0.0485 & 0.69 & 0.0727 & 1.35 & 0.1350 \\
   \bottomrule
  \end{tabular}
\end{table*}

\begin{table}[t]
  \caption{EER[\%] of VoxCeleb 1-O of different adaptation methods applied on NEMO Small ASR-CTC model (15.88M). \tablefootnote{\#ASR param indicates the model size (in million parameters) of the ASR Conformer encoder when it has $L$ layers. \#adap param represents the  \textcolor{black}{speaker adaptation modules'} total model size including $L$ adaptor layers (in million parameters).}}
  \label{tab:result3}
  \centering
  \begin{tabular}[c]{@{\ }c@{\ \ \ }c@{\ }|@{\ \ }c@{\ \ \ \ }c@{\ \ }|@{\ \ }c@{\ \ \ \ }c@{\ \ }|@{\ \ }c@{\ \ \ \ }c@{\ \ }|@{\ \ }c@{\ }}
    \toprule
    & \multicolumn{2}{c}{$L$}  & \multicolumn{2}{c}{$K=0$} & \multicolumn{2}{c}{$K=2$} & \multicolumn{2}{c}{$K=4$} \\
    & \#$L$ & \makecell{\#ASR\\param} & EER & \makecell{\#adap\\param} & EER & \makecell{\#adap\\param} & EER & \makecell{\#adap\\param} \\
    \midrule
    \multirow{3}*{V1} & 4 & 3.92M & 2.49 & 0.73M & 1.22 & 2.60M & 1.21 & 4.47M \\
     & 8 & 6.94M & 1.77 & 1.45M & 1.26 & 3.32M & 1.12 & 5.20M \\
     & 12 & 9.95M & 1.73 & 2.18M & 1.47 & 4.05M & 1.34 & 5.92M \\
    \midrule
    \multirow{3}*{V2} & 4 & 3.92M & 1.47 & 0.69M & 1.13 & 2.56M & 1.05 & 4.43M \\
     & 8 & 6.94M & 1.11 & 1.37M & 1.02 & 3.24M & 0.94 & 5.12M \\
     & 12 & 9.95M & 1.10 & 2.06M & 1.03 & 3.93M & 1.03 & 5.80M \\
    \midrule
    \multirow{3}*{V3} & 4 & 3.92M & \multicolumn{2}{c}{---} & 0.98 & 2.68M & 0.95 & 4.55M \\
     & 8 & 6.94M & \multicolumn{2}{c}{---} & 0.83 & 3.49M & 0.83 & 5.36M \\
     & 12 & 9.95M & \multicolumn{2}{c}{---} & 0.79 & 4.30M & 0.66 & 6.17M \\
    \bottomrule
  \end{tabular}
\end{table}

\begin{table}[t]
  \caption{EER[\%] of VoxCeleb 1-O of different adaptation methods applied on NEMO Medium ASR-CTC model (35.26M).$^\text{7}$}
  \label{tab:result4}
  \centering
  \begin{tabular}[c]{@{\ }c@{\ \ \ }c@{\ }|@{\ \ }c@{\ \ \ \ }c@{\ \ }|@{\ \ }c@{\ \ \ \ }c@{\ \ }|@{\ \ }c@{\ \ \ \ }c@{\ \ }|@{\ \ }c@{\ }}
    \toprule
    & \multicolumn{2}{c}{$L$}  & \multicolumn{2}{c}{$K=0$} & \multicolumn{2}{c}{$K=2$} & \multicolumn{2}{c}{$K=4$} \\
    & \#$L$ & \makecell{\#ASR\\param} & EER & \makecell{\#adap\\param} & EER & \makecell{\#adap\\param} & EER & \makecell{\#adap\\param} \\
    \midrule
    \multirow{3}*{V1} & 6 & 11.44M & 1.65 & 1.63M & 1.01 & 3.50M & 0.99 & 5.37M \\
     & 10 & 17.79M & 1.40 & 2.69M & 1.10 & 4.56M & 1.03 & 6.43M \\
     & 14 & 24.15M & 1.34 & 3.74M & 1.20 & 5.61M & 1.19 & 7.48M \\
    \midrule
    \multirow{3}*{V2} & 6 & 11.44M & 1.08 & 1.14M & 0.89 & 3.01M & 0.94 & 4.88M \\
     & 10 & 17.79M & 0.93 & 2.69M & 0.84 & 4.56M & 0.84 & 6.43M \\
     & 14 & 24.15M & 0.89 & 3.74M & 0.92 & 5.61M & 0.86 & 7.48M \\
    \midrule
    \multirow{3}*{V3} & 6 & 11.44M & \multicolumn{2}{c}{---} & 0.81 & 3.23M & 0.90 & 5.10M \\
     & 10 & 17.79M & \multicolumn{2}{c}{---} & 0.67 & 4.14M & 0.83 & 6.01M \\
     & 14 & 24.15M & \multicolumn{2}{c}{---} & 0.66 & 5.05M & 0.77 & 6.92M \\
    \bottomrule
  \end{tabular}
\end{table}

\begin{table}[t]
  \caption{EER[\%] of VoxCeleb 1-O of different adaptation methods applied on NEMO Large ASR-CTC model (130.94M).$^\text{7}$}
  \label{tab:result5}
  \centering
  \begin{tabular}[c]{@{\ }c@{\ \ \ }c@{\ }|@{\ \ }c@{\ \ \ \ }c@{\ \ }|@{\ \ }c@{\ \ \ \ }c@{\ \ }|@{\ \ }c@{\ \ \ \ }c@{\ \ }|@{\ \ }c@{\ }}
    \toprule
    & \multicolumn{2}{c}{$L$}  & \multicolumn{2}{c}{$K=0$} & \multicolumn{2}{c}{$K=2$} & \multicolumn{2}{c}{$K=4$} \\
    & \#$L$ & \makecell{\#ASR\\param} & EER & \makecell{\#adap\\param} & EER & \makecell{\#adap\\param} & EER & \makecell{\#adap\\param} \\
    \midrule
    \multirow{3}*{V1} & 6 & 45.55M & 1.18 & 3.26M & 0.88 & 5.13M & 0.94 & 7.00M \\
     & 10 & 70.85M & 0.97 & 5.37M & 0.85 & 7.24M & 0.89 & 9.11M \\
     & 14 & 96.14M & 1.01 & 7.48M & 0.89 & 9.35M & 0.97 & 11.23M \\
    \midrule
    \multirow{3}*{V2} & 6 & 45.55M & 0.86 & 1.38M & 0.72 & 3.25M & 0.78 & 5.12M \\
     & 10 & 70.85M & 0.72 & 2.23M & 0.78 & 4.11M & 0.75 & 5.98M \\
     & 14 & 96.14M & 0.71 & 3.09M & 0.77 & 4.96M & 0.76 & 6.84M \\
    \midrule
    \multirow{3}*{V3} & 6 & 45.55M & \multicolumn{2}{c}{---} & 0.61 & 3.70M & 0.69 & 5.57M \\
     & 10 & 70.85M & \multicolumn{2}{c}{---} & 0.57 & 4.92M & 0.65 & 6.79M \\
     & 14 & 96.14M & \multicolumn{2}{c}{---} & 0.55 & 6.14M & 0.65 & 8.01M \\
    \bottomrule
  \end{tabular}
\end{table}

\subsection{Speaker adaptation: unifying ASR and speaker verification}\label{exp3}
In this section, we delve into the effectiveness of our proposed speaker adaptation approach in bridging the gap between ASR and speaker verification tasks.

We employ three pretrained ASR Conformer encoders — small, medium, and large, as referenced in Table \ref{tab:netarc}.
These encoders are integrated with our speaker adaptation technique.
For each encoder size, we assess three distinct configurations of the  \textcolor{black}{speaker adaptation module}, based on the one depicted in Fig. \ref{fig: adaptor}:
\begin{itemize}
	\item \textbf{V1}: This version extracts directly from the first $L$ layers of the ASR Conformer without the intervention of layer adaptors.
	\item \textbf{V2}: In alignment with Fig. \ref{fig: adaptor a}, this version integrates layer adaptors to refine the outputs from the first $L$ ASR Conformer layers. Subsequently, $K$ lightweight Conformer layers process the frame-level outputs derived from the $L$-th ASR Conformer layer.
	\item \textbf{V3}: As illustrated in Fig. \ref{fig: adaptor b}, this configuration feeds the $K$ lightweight Conformer layers with a concatenated output from the first $L$ Conformer layers of the pretrained ASR model. An auxiliary linear layer ensures the alignment of concatenated feature dimensions.
\end{itemize}
All configurations use a lightweight Conformer layer architecture consistent across the ASR encoders.
These lightweight Conformer layers have 174 dimensions, 704 hidden units, and 4 attention heads, the same as the Conformer layer configuration in the NEMO Small ASR-CTC model.
Additionally, the layer adaptor always maps the frame-level outputs from ASR Conformer layers to a 128-dimensional feature space.
A detailed architecture configuration can be found in Table \ref{tab: adapter arch}.
Notably, when the model lacks trainable Conformer layers (i.e., $K=0$), the V2 and V3 configurations converge to become identical.
The specific EERs for distinct model configurations, considering variations in both $L$ and $K$, are outlined in Tables \ref{tab:result3}, \ref{tab:result4}, and \ref{tab:result5}, each corresponding to a unique pretrained ASR model.

\subsubsection{Baseline - ASR Conformer without speaker adaptation}
Before introducing any adaptation method, it is crucial to understand the innate capabilities of the ASR Conformer encoder when used for speaker verification.
Our baseline is free from any layer adaptor (configuration V1) and does not incorporate additional trainable Conformer layers ($K=0$).
 Here, the frame-level outputs of the ASR Conformer are concatenated and subsequently routed to the pooling layer to extract speaker embeddings.
The results consistently indicate a notable trend: ASR models with a more significant number of parameters (or layers) often exhibit superior performance compared to their smaller counterparts.
For instance, while the NEMO Small ASR-CTC model with 12 layers has an EER of 1.73\%, its larger counterpart, the NEMO Large ASR-CTC model with 6 layers, surpasses it with a more desirable EER of 1.18\%.
While increasing the ASR Conformer's layers generally leads to a decrease in the EER, the relationship is not strictly linear.
For instance, in NEMO Large ASR-CTC mode, while moving from 6 to 10 layers results in an EER reduction from 1.18\% to 0.97\%, further increasing to 14 layers sees a slight EER increase to 1.01\%.

\subsubsection{ASR Conformer with layer adaptors}
After assessing the ASR Conformer without speaker adaptation, we investigated the effect of introducing layer adaptors (configuration V2) without integrating additional trainable Conformer layers ($K=0$).
Using the layer adaptor, the ASR Conformer's feature dimensions are reduced to 128, resulting in a smaller concatenated feature dimension after MFA concatenation.
This led to a more compact  \textcolor{black}{speaker adaptation module} in V2 compared to V1.
Our findings indicate that introducing layer adaptors substantially enhances the speaker verification performance.
Specifically, for the NEMO Small ASR-CTC model with $L=12$, we observed an EER of 1.10\%, marking a relative 36\% reduction from the baseline's 1.73\% in the absence of speaker adaptation.
Similar performance improvements are also witnessed across medium and large ASR-CTC models.
The consistent performance improvement across different model sizes proves the effectiveness of layer adaptors.

\subsubsection{ASR Conformer with trainable lightweight Conformer layers}
Expanding our investigation, we delved into the impact of incorporating trainable lightweight Conformer layers into the ASR Conformer under configuration V1, specifically with $K=2$ and $K=4$.
Adding additional trainable layers to the ASR Conformer resulted in improved performance.
Compared to the baseline model, adding just two trainable layers demonstrated a marked reduction in EER across all configurations.
However, these performance gains tend to plateau. For instance, while adding 2 trainable layers yields a noteworthy improvement, the benefits diminish, or in some cases even slightly reverse, with the addition of 4 layers.
One plausible explanation is that the inputs to these lightweight trainable Conformer layers come from highly abstract signals from the ASR model. Therefore, an increase in their number could potentially lead to overfitting.

\subsubsection{Comparing the input of the trainable Conformer layers}
Our subsequent investigation aimed at the inputs channeled into the trainable lightweight Conformer layers.
We compared configurations V2 and V3, explicitly focusing on $K=2$ and $K=4$.
In configuration V2, the inputs to the trainable Conformer layer are sourced directly from the frame-level outputs derived from the $L$-th ASR Conformer layer.
Conversely, in configuration V3, the trainable Conformer layer receives its inputs from a concatenation sourced from the ASR model's first $L$ Conformer layers.
A clear distinction in performance emerged from the results: Configuration V3 consistently outperforms V2 across all ASR model sizes and all values of $L$.
For instance, considering the NEMO Large ASR-CTC model with $L=14$ and $K=2$, V3 achieved an EER of 0.55\%, this translates to a relative reduction of 29\% compared to V2.
As shown in the linear probe experiments in section \ref{sec: adaptor}, the early layers of the ASR Conformer model are proficient at gathering speaker-specific information.
The concatenation from multiple ASR Conformer layers in V3 captures a more diverse and quality-rich set of information, which proves advantageous for the  \textcolor{black}{speaker adaptation module}.

For a more thorough evaluation, we test the speaker adaptation method on three testing trials of VoxCeleb 1. We select one  \textcolor{black}{speaker adaptation module} with the V3 configuration for each NEMO ASR Conformer-CTC model of varying sizes. The results can be found in Table \ref{tab:result6}.
The V3  \textcolor{black}{speaker adaptation module} with $L=10$ and $K=2$ achieves an 0.45\% EER using the NEMO Large ASR-CTC model.
In comparison, the ASR pretraining and ASR distillation techniques result in EERs of 0.43\% and 0.45\%, respectively, using the same Large model.
While the speaker adaptation method lags slightly behind these two methods, it uniquely offers the capability of unifying ASR and speaker verification within a single Conformer model.
This benefit of task unification comes with a relatively modest increase of 4.92 million parameters added to the 130.94 million parameter Large ASR Conformer encoder.

\section{Conclusion}
This research has presented and evaluated three techniques to leverage ASR pretrained Conformers for speaker verification tasks effectively. Experiments on VoxCeleb datasets validate the efficacy of our proposed methods.
First, we have shown that initializing speaker embedding networks with ASR pretrained Conformers lead to significant performance gains and generalization. The extensive ASR pretraining enables the network to extract more robust speaker representations by preventing overfitting to limited speaker data.
Second, knowledge distillation from the ASR Conformer teacher to the speaker verification student model allows efficient transfer of ASR expertise. Serving as an auxiliary phonetic modeling task, this distillation approach enhances speaker modeling. Compared to direct ASR pretraining, knowledge distillation offers more flexibility in student model design.
Third, our lightweight adaptation modules successfully unify ASR and speaker verification within a single Conformer model. By refining ASR-learned features for speaker tasks, the adaptation module efficiently bridges the gap between the two modalities. This unified model delivers simultaneous ASR and speaker verification using minimal additional parameters.
This research has demonstrated three promising and viable strategies to leverage ASR pretrained Conformers to advance speaker verification performance. Our methods effectively transfer rich ASR knowledge to speaker modeling. We aim to extend our approaches to multilingual models and low-resource settings for further studies.
\bibliographystyle{IEEEtran}
\bibliography{refs}

\begin{thebibliography}{10}
\providecommand{\url}[1]{#1}
\csname url@samestyle\endcsname
\providecommand{\newblock}{\relax}
\providecommand{\bibinfo}[2]{#2}
\providecommand{\BIBentrySTDinterwordspacing}{\spaceskip=0pt\relax}
\providecommand{\BIBentryALTinterwordstretchfactor}{4}
\providecommand{\BIBentryALTinterwordspacing}{\spaceskip=\fontdimen2\font plus
\BIBentryALTinterwordstretchfactor\fontdimen3\font minus \fontdimen4\font\relax}
\providecommand{\BIBforeignlanguage}[2]{{%
\expandafter\ifx\csname l@#1\endcsname\relax
\typeout{** WARNING: IEEEtran.bst: No hyphenation pattern has been}%
\typeout{** loaded for the language `#1'. Using the pattern for}%
\typeout{** the default language instead.}%
\else
\language=\csname l@#1\endcsname
\fi
#2}}
\providecommand{\BIBdecl}{\relax}
\BIBdecl

\bibitem{snyder_x-vectors:_2018}
D.~Snyder, D.~{Garcia-Romero}, G.~Sell, D.~Povey, and S.~Khudanpur, ``X-vectors: {{Robust DNN Embeddings}} for {{Speaker Recognition}},'' in \emph{ICASSP}, 2018, pp. 5329--5333.

\bibitem{cai_--fly_2020}
W.~Cai, J.~Chen, J.~Zhang, and M.~Li, ``On-the-{{Fly Data Loader}} and {{Utterance-Level Aggregation}} for {{Speaker}} and {{Language Recognition}},'' \emph{IEEE/ACM TASLP}, vol.~28, pp. 1038--1051, 2020.

\bibitem{cai_exploring_2018}
W.~Cai, J.~Chen, and M.~Li, ``Exploring the {Encoding} {Layer} and {Loss} {Function} in {End}-to-{End} {Speaker} and {Language} {Recognition} {System},'' in \emph{Speaker Odyssey}, 2018, pp. 74--81.

\bibitem{okabe_attentive_2018}
K.~Okabe, T.~Koshinaka, and K.~Shinoda, ``Attentive {{Statistics Pooling}} for {{Deep Speaker Embedding}},'' in \emph{Interspeech}, 2018, pp. 2252--2256.

\bibitem{zhou_resnext_2021}
T.~Zhou, Y.~Zhao, and J.~Wu, ``{{ResNeXt}} and {{Res2Net Structures}} for {{Speaker Verification}},'' in \emph{SLT}, 2021, pp. 301--307.

\bibitem{desplanques_ecapa-tdnn_2020}
B.~Desplanques, J.~Thienpondt, and K.~Demuynck, ``{{ECAPA-TDNN}}: {{Emphasized Channel Attention}}, {{Propagation}} and {{Aggregation}} in {{TDNN Based Speaker Verification}},'' in \emph{Interspeech}, 2020, pp. 3830--3834.

\bibitem{deng_arcface_2019}
J.~Deng, J.~Guo, N.~Xue, and S.~Zafeiriou, ``{{ArcFace}}: {{Additive Angular Margin Loss}} for {{Deep Face Recognition}},'' in \emph{{{CVPR}}}, 2019, pp. 4685--4694.

\bibitem{9023039}
X.~Xiang, S.~Wang, H.~Huang, Y.~Qian, and K.~Yu, ``{Margin Matters: Towards More Discriminative Deep Neural Network Embeddings for Speaker Recognition},'' in \emph{APSIPA}, 2019, pp. 1652--1656.

\bibitem{chung2020in}
J.~S. Chung, J.~Huh, S.~Mun, M.~Lee, H.~S. Heo, S.~Choe, C.~Ham, S.~Jung, B.-J. Lee, and I.~Han, ``{In Defence of Metric Learning for Speaker Recognition},'' in \emph{Interspeech}, 2020, pp. 2977--2981.

\bibitem{garcia-romero_magneto_2020}
D.~{Garcia-Romero}, G.~Sell, and A.~Mccree, ``{{MagNetO}}: {{X-vector Magnitude Estimation Network}} plus {{Offset}} for {{Improved Speaker Recognition}},'' in \emph{Odyssey}, 2020, pp. 1--8.

\bibitem{9414600}
J.~Thienpondt, B.~Desplanques, and K.~Demuynck, ``{The Idlab Voxsrc-20 Submission: Large Margin Fine-Tuning and Quality-Aware Score Calibration in DNN Based Speaker Verification},'' in \emph{ICASSP}, 2021, pp. 5814--5818.

\bibitem{he_deep_2016}
K.~He, X.~Zhang, S.~Ren, and J.~Sun, ``Deep {{Residual Learning}} for {{Image Recognition}},'' in \emph{{{CVPR}}}, 2016, pp. 770--778.

\bibitem{Hu_2018_CVPR}
J.~Hu, L.~Shen, and G.~Sun, ``{Squeeze-and-Excitation Networks},'' in \emph{CVPR}, 2018.

\bibitem{8821313}
S.-H. Gao, M.-M. Cheng, K.~Zhao, X.-Y. Zhang, M.-H. Yang, and P.~Torr, ``{Res2Net: A New Multi-Scale Backbone Architecture},'' \emph{IEEE TPAMI}, vol.~43, no.~2, pp. 652--662, 2021.

\bibitem{Xie_2017_CVPR}
S.~Xie, R.~Girshick, P.~Dollar, Z.~Tu, and K.~He, ``{Aggregated Residual Transformations for Deep Neural Networks},'' in \emph{CVPR}, 2017.

\bibitem{vaswani2017attention}
A.~Vaswani, N.~Shazeer, N.~Parmar, J.~Uszkoreit, L.~Jones, A.~N. Gomez, {\L}.~Kaiser, and I.~Polosukhin, ``{Attention is All You Need},'' in \emph{NeurIPS}, 2017.

\bibitem{gulati_conformer_2020}
A.~Gulati, J.~Qin, C.-C. Chiu, N.~Parmar, Y.~Zhang, J.~Yu, W.~Han, S.~Wang, Z.~Zhang, Y.~Wu, and R.~Pang, ``Conformer: {{Convolution-augmented Transformer}} for {{Speech Recognition}},'' in \emph{Interspeech}, 2020, pp. 5036--5040.

\bibitem{zhang_mfa-conformer_2022}
Y.~Zhang, Z.~Lv, H.~Wu, S.~Zhang, P.~Hu, Z.~Wu, H.-y. Lee, and H.~Meng, ``{{MFA-Conformer}}: {{Multi-scale Feature Aggregation Conformer}} for {{Automatic Speaker Verification}},'' in \emph{Interspeech}, 2022, pp. 306--310.

\bibitem{10095433}
D.~Liao, T.~Jiang, F.~Wang, L.~Li, and Q.~Hong, ``{Towards A Unified Conformer Structure: from ASR to ASV Task},'' in \emph{ICASSP}, 2023, pp. 1--5.

\bibitem{10096659}
D.~Cai, W.~Wang, M.~Li, R.~Xia, and C.~Huang, ``{Pretraining Conformer with ASR for Speaker Verification},'' in \emph{ICASSP}, 2023, pp. 1--5.

\bibitem{zhou_cnn_2019}
T.~Zhou, Y.~Zhao, J.~Li, Y.~Gong, and J.~Wu, ``{{CNN}} with {{Phonetic Attention}} for {{Text-Independent Speaker Verification}},'' in \emph{ASRU}, 2019, pp. 718--725.

\bibitem{li2016generalized}
M.~Li, L.~Liu, W.~Cai, and W.~Liu, ``{Generalized I-vector Representation with Phonetic Tokenizations and Tandem Features for both Text Independent and Text Dependent Speaker Verification},'' \emph{Journal of Signal Processing Systems}, vol.~82, no.~2, pp. 207--215, 2016.

\bibitem{6853887}
Y.~Lei, N.~Scheffer, L.~Ferrer, and M.~McLaren, ``{A Novel Scheme for Speaker Recognition using a Phonetically-Aware Deep Neural Network},'' in \emph{ICASSP}, 2014, pp. 1695--1699.

\bibitem{hinton_distilling_2015}
G.~Hinton, O.~Vinyals, and J.~Dean, ``Distilling the {{Knowledge}} in a {{Neural Network}},'' in \emph{{{NeurIPS Deep Learning}} and {{Representation Learning Workshop}}}, 2015.

\bibitem{fan_exploring_2021}
Z.~Fan, M.~Li, S.~Zhou, and B.~Xu, ``{Exploring Wav2vec 2.0 on Speaker Verification and Language Identification},'' in \emph{Interspeech}, 2021, pp. 1509--1513.

\bibitem{vaessen_fine-tuning_2022}
N.~Vaessen and D.~A. {van Leeuwen}, ``{Fine-Tuning Wav2vec2 for Speaker Recognition},'' in \emph{{{ICASSP}}}, 2022, pp. 7967--7971.

\bibitem{novoselov_robust_2022}
S.~Novoselov, G.~Lavrentyeva, A.~Avdeeva, V.~Volokhov, and A.~Gusev, ``Robust {{Speaker Recognition}} with {{Transformers Using}} wav2vec 2.0,'' \emph{arXiv:2203.15095}, 2022.

\bibitem{chen_large-scale_2022}
Z.~Chen, S.~Chen, Y.~Wu, Y.~Qian, C.~Wang, S.~Liu, Y.~Qian, and M.~Zeng, ``{Large-Scale} {{Self-Supervised Speech Representation Learning}} for {{Automatic Speaker Verification}},'' in \emph{{{ICASSP}}}, 2022, pp. 6147--6151.

\bibitem{9814838}
S.~Chen, C.~Wang, Z.~Chen, Y.~Wu, S.~Liu, Z.~Chen, J.~Li, N.~Kanda, T.~Yoshioka, X.~Xiao, J.~Wu, L.~Zhou, S.~Ren, Y.~Qian, Y.~Qian, J.~Wu, M.~Zeng, X.~Yu, and F.~Wei, ``{WavLM: Large-Scale Self-Supervised Pre-Training for Full Stack Speech Processing},'' \emph{IEEE Journal of Selected Topics in Signal Processing}, vol.~16, no.~6, pp. 1505--1518, 2022.

\bibitem{peng_attention-based_2022}
J.~Peng, O.~Plchot, T.~Stafylakis, L.~Mo{\v s}ner, L.~Burget, and J.~{\v C}ernock{\'y}, ``An {{Attention-Based Backend Allowing Efficient Fine-Tuning}} of {{Transformer Models}} for {{Speaker Verification}},'' in \emph{{{SLT}}}, 2022, pp. 555--562.

\bibitem{7404779}
D.~Snyder, D.~Garcia-Romero, and D.~Povey, ``{Time Delay Deep Neural Network-based Universal Background Models for Speaker Recognition},'' in \emph{ASRU}, 2015, pp. 92--97.

\bibitem{tian15_interspeech}
Y.~Tian, M.~Cai, L.~He, and J.~Liu, ``{Investigation of Bottleneck Features and Multilingual Deep Neural Networks for Speaker Verification},'' in \emph{Interspeech}, 2015, pp. 1151--1155.

\bibitem{rahman_employing_2018}
M.~H. Rahman, I.~Himawan, M.~McLaren, C.~Fookes, and S.~Sridharan, ``Employing {{Phonetic Information}} in {{DNN Speaker Embeddings}} to {{Improve Speaker Recognition Performance}},'' in \emph{Interspeech}, 2018, pp. 3593--3597.

\bibitem{zheng_phonetically-aware_2020}
S.~Zheng, Y.~Lei, and H.~Suo, ``Phonetically-{{Aware Coupled Network For Short Duration Text-Independent Speaker Verification}},'' in \emph{Interspeech}, 2020, pp. 926--930.

\bibitem{liu_speaker_2018}
Y.~Liu, L.~He, J.~Liu, and M.~T. Johnson, ``Speaker {{Embedding Extraction}} with {{Phonetic Information}},'' in \emph{Interspeech}, 2018, pp. 2247--2251.

\bibitem{chen_phoneme-unit-specific_2021}
X.~Chen and C.~Bao, ``Phoneme-{{Unit-Specific Time-Delay Neural Network}} for {{Speaker Verification}},'' \emph{IEEE/ACM TASLP}, vol.~29, pp. 1243--1255, 2021.

\bibitem{tang_collaborative_2017}
Z.~Tang, L.~Li, D.~Wang, and R.~Vipperla, ``Collaborative {{Joint Training With Multitask Recurrent Model}} for {{Speech}} and {{Speaker Recognition}},'' \emph{IEEE/ACM TASLP}, vol.~25, no.~3, pp. 493--504, 2017.

\bibitem{wang_usage_2019}
S.~Wang, J.~Rohdin, L.~Burget, O.~Plchot, Y.~Qian, K.~Yu, and J.~{\v C}ernock{\'y}, ``On the {{Usage}} of {{Phonetic Information}} for {{Text-Independent Speaker Embedding Extraction}},'' in \emph{Interspeech}, 2019, pp. 1148--1152.

\bibitem{tawara_frame-level_2020-1}
N.~Tawara, A.~Ogawa, T.~Iwata, M.~Delcroix, and T.~Ogawa, ``Frame-{{Level Phoneme-Invariant Speaker Embedding}} for {{Text-Independent Speaker Recognition}} on {{Extremely Short Utterances}},'' in \emph{{{ICASSP}}}, 2020, pp. 6799--6803.

\bibitem{hong_decomposition_2023}
Q.-B. Hong, C.-H. Wu, and H.-M. Wang, ``Decomposition and {{Reorganization}} of {{Phonetic Information}} for {{Speaker Embedding Learning}},'' \emph{IEEE/ACM TASLP}, vol.~31, pp. 1745--1757, 2023.

\bibitem{houlsby_parameter-efficient_2019}
N.~Houlsby, A.~Giurgiu, S.~Jastrzebski, B.~Morrone, Q.~{de Laroussilhe}, A.~Gesmundo, M.~Attariyan, and S.~Gelly, ``Parameter-{{Efficient Transfer Learning}} for {{NLP}},'' in \emph{{{ICML}}}, 2019, pp. 2790--2799.

\bibitem{rebuffi2017learning}
S.-A. Rebuffi, H.~Bilen, and A.~Vedaldi, ``{Learning Multiple Visual Domains with Residual Adapters},'' in \emph{NeurIPS}, vol.~30, 2017.

\bibitem{he_towards_2022}
J.~He, C.~Zhou, X.~Ma, T.~{Berg-Kirkpatrick}, and G.~Neubig, ``Towards a {{Unified View}} of {{Parameter-Efficient Transfer Learning}},'' in \emph{{{ICLR}}}, 2022.

\bibitem{bapna_simple_2019}
A.~Bapna, N.~Arivazhagan, and O.~Firat, ``Simple, {{Scalable Adaptation}} for {{Neural Machine Translation}},'' in \emph{{{EMNLP}}}, 2019.

\bibitem{thomas_efficient_2022-1}
B.~Thomas, S.~Kessler, and S.~Karout, ``Efficient {{Adapter Transfer}} of {{Self-Supervised Speech Models}} for {{Automatic Speech Recognition}},'' in \emph{{{ICASSP}}}, 2022, pp. 7102--7106.

\bibitem{kannan_large-scale_2019}
A.~Kannan, A.~Datta, T.~N. Sainath, E.~Weinstein, B.~Ramabhadran, Y.~Wu, A.~Bapna, Z.~Chen, and S.~Lee, ``Large-{{Scale Multilingual Speech Recognition}} with a {{Streaming End-to-End Model}},'' in \emph{Interspeech}, 2019, pp. 2130--2134.

\bibitem{winata_adapt-and-adjust_2021}
G.~I. Winata, G.~Wang, C.~Xiong, and S.~Hoi, ``Adapt-and-{{Adjust}}: {{Overcoming}} the {{Long-Tail Problem}} of {{Multilingual Speech Recognition}},'' in \emph{Interspeech}, 2021, pp. 2451--2455.

\bibitem{le_lightweight_2021}
H.~Le, J.~Pino, C.~Wang, J.~Gu, D.~Schwab, and L.~Besacier, ``Lightweight {{Adapter Tuning}} for {{Multilingual Speech Translation}},'' in \emph{{{ACL-IJCNLP}}}, 2021, pp. 817--824.

\bibitem{peng_parameter-efficient_2023}
J.~Peng, T.~Stafylakis, R.~Gu, O.~Plchot, L.~Mo{\v s}ner, L.~Burget, and J.~{\v C}ernock{\'y}, ``Parameter-{{Efficient Transfer Learning}} of {{Pre-Trained Transformer Models}} for {{Speaker Verification Using Adapters}},'' in \emph{{{ICASSP}}}, 2023, pp. 1--5.

\bibitem{otake_parameter_2023}
S.~Otake, R.~Kawakami, and N.~Inoue, ``Parameter {{Efficient Transfer Learning}} for {{Various Speech Processing Tasks}},'' in \emph{{{ICASSP}}}, 2023, pp. 1--5.

\bibitem{dai2019transformer}
Z.~Dai, Z.~Yang, Y.~Yang, J.~G. Carbonell, Q.~Le, and R.~Salakhutdinov, ``{Transformer-XL: Attentive Language Models beyond a Fixed-Length Context},'' in \emph{ACL}, 2019, pp. 2978--2988.

\bibitem{lu2019understanding}
Y.~Lu, Z.~Li, D.~He, Z.~Sun, B.~Dong, T.~Qin, L.~Wang, and T.-Y. Liu, ``{Understanding and Improving Transformer from a Multi-Particle Dynamic System Point of View},'' \emph{arXiv:1906.02762}, 2019.

\bibitem{okabe18_interspeech}
K.~Okabe, T.~Koshinaka, and K.~Shinoda, ``{Attentive Statistics Pooling for Deep Speaker Embedding},'' in \emph{Interspeech}, 2018, pp. 2252--2256.

\bibitem{hsu_hubert_2021-1}
W.-N. Hsu, B.~Bolte, Y.-H.~H. Tsai, K.~Lakhotia, R.~Salakhutdinov, and A.~Mohamed, ``{{HuBERT}}: {{Self-Supervised Speech Representation Learning}} by {{Masked Prediction}} of {{Hidden Units}},'' \emph{IEEE/ACM TASLP}, vol.~29, pp. 3451--3460, 2021.

\bibitem{xu-etal-2021-optimizing}
P.~Xu, D.~Kumar, W.~Yang, W.~Zi, K.~Tang, C.~Huang, J.~C.~K. Cheung, S.~J. Prince, and Y.~Cao, ``{Optimizing Deeper Transformers on Small Datasets},'' in \emph{ACL IJCNLP}, 2021, pp. 2089--2102.

\bibitem{nagrani_voxceleb:_2017}
A.~Nagrani, J.~S. Chung, and A.~Zisserman, ``Voxceleb: {A} {Large}-{Scale} {Speaker} {Identification} {Dataset},'' in \emph{Interspeech}, 2017, pp. 2616--2620.

\bibitem{chung_voxceleb2:_2018}
J.~S. Chung, A.~Nagrani, and A.~Zisserman, ``Voxceleb2: {{Deep Speaker Recognition}},'' in \emph{Interspeech}, 2018, pp. 1086--1090.

\bibitem{yamamoto19_interspeech}
H.~Yamamoto, K.~A. Lee, K.~Okabe, and T.~Koshinaka, ``{Speaker Augmentation and Bandwidth Extension for Deep Speaker Embedding},'' in \emph{Interspeech}, 2019, pp. 406--410.

\bibitem{wang2020dku}
W.~Wang, D.~Cai, X.~Qin, and M.~Li, ``{The DKU-DukeECE Systems for VoxCeleb Speaker Recognition Challenge 2020},'' \emph{arXiv:2010.12731}, 2020.

\bibitem{musan}
D.~Snyder, G.~Chen, and D.~Povey, ``{MUSAN}: {A} {Music}, {Speech}, and {Noise} {Corpus},'' \emph{arXiv:1510.08484}, 2015.

\bibitem{ko2017study}
T.~Ko, V.~Peddinti, D.~Povey, M.~L. Seltzer, and S.~Khudanpur, ``{A Study on Data Augmentation of Reverberant Speech for Robust Speech Recognition},'' in \emph{ICASSP}, 2017, pp. 5220--5224.

\bibitem{kuchaiev2019nemo}
O.~Kuchaiev, J.~Li, H.~Nguyen, O.~Hrinchuk, R.~Leary, B.~Ginsburg, S.~Kriman, S.~Beliaev, V.~Lavrukhin, J.~Cook \emph{et~al.}, ``{Nemo: a Toolkit for Building AI Applications Using Neural Modules},'' \emph{arXiv:1909.09577}, 2019.

\bibitem{sn_analysis_2017}
P.~Matějka, O.~Novotný, O.~Plchot, L.~Burget, M.~D. Sánchez, and J.~Černocký, ``Analysis of {Score} {Normalization} in {Multilingual} {Speaker} {Recognition},'' in \emph{Interspeech}, 2017, pp. 1567--1571.

\bibitem{6584746}
M.~I. Mandasari, R.~Saeidi, M.~McLaren, and D.~A. van Leeuwen, ``{Quality Measure Functions for Calibration of Speaker Recognition Systems in Various Duration Conditions},'' \emph{IEEE TASLP}, vol.~21, no.~11, pp. 2425--2438, 2013.

\bibitem{nist_nist_2016}
\BIBentryALTinterwordspacing
``{{NIST}} 2016 {{Speaker Recognition Evaluation Plan}},'' 2016. [Online]. Available: \url{https://www.nist.gov/system/files/documents/2016/10/07/sre16_eval_plan_v1.3.pdf}
\BIBentrySTDinterwordspacing

\end{thebibliography}

\begin{IEEEbiography}[{\includegraphics[width=1in,height=1.25in,clip,keepaspectratio]{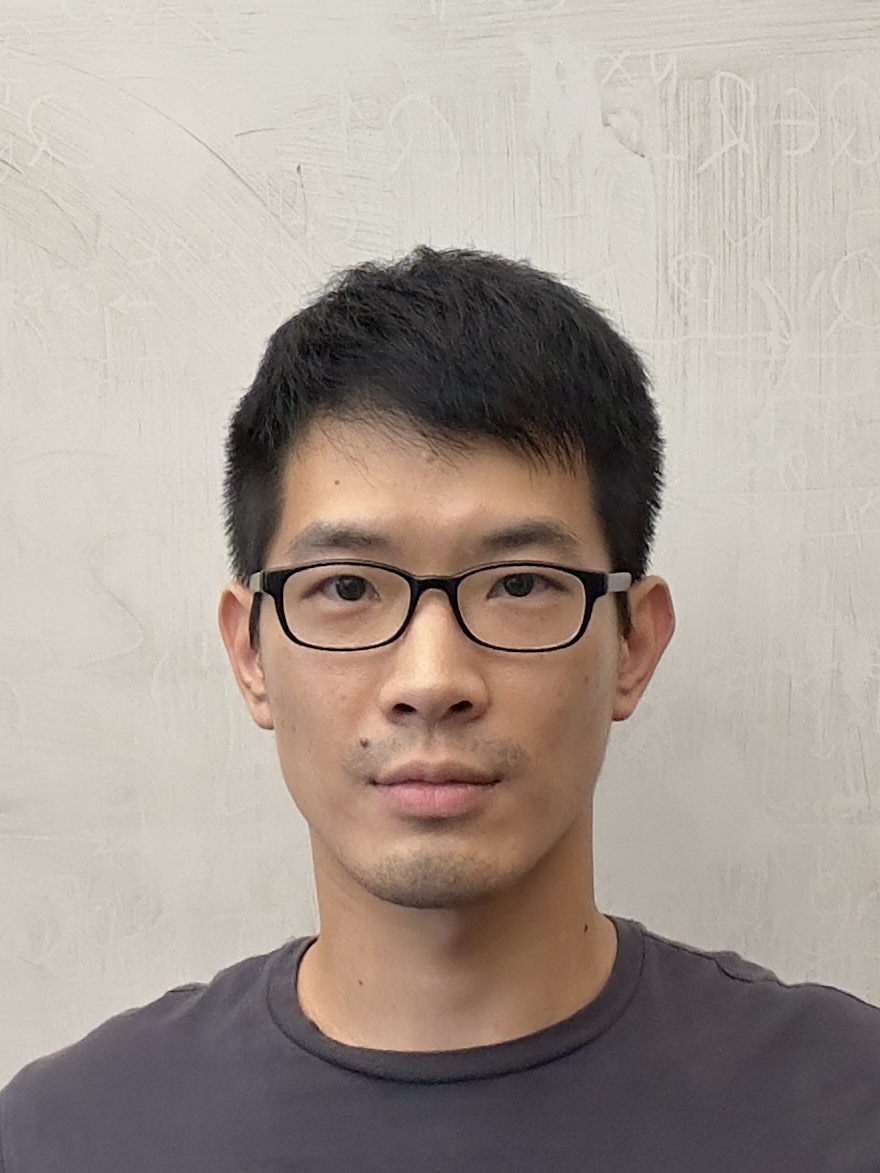}}]{Danwei Cai}
is pursuing his Ph.D. degree in electrical and computer engineering at Duke University. He received his bachelor's degree in software engineering and master's degree in electronics and communication engineering from Sun Yet-Sen University in China. His primary research interests are in the area of speech processing, including speech recognition, speaker recognition, speaker diarization and computational linguistics.
\end{IEEEbiography}
\begin{IEEEbiography}[{\includegraphics[width=1in,height=1.25in,clip,keepaspectratio]{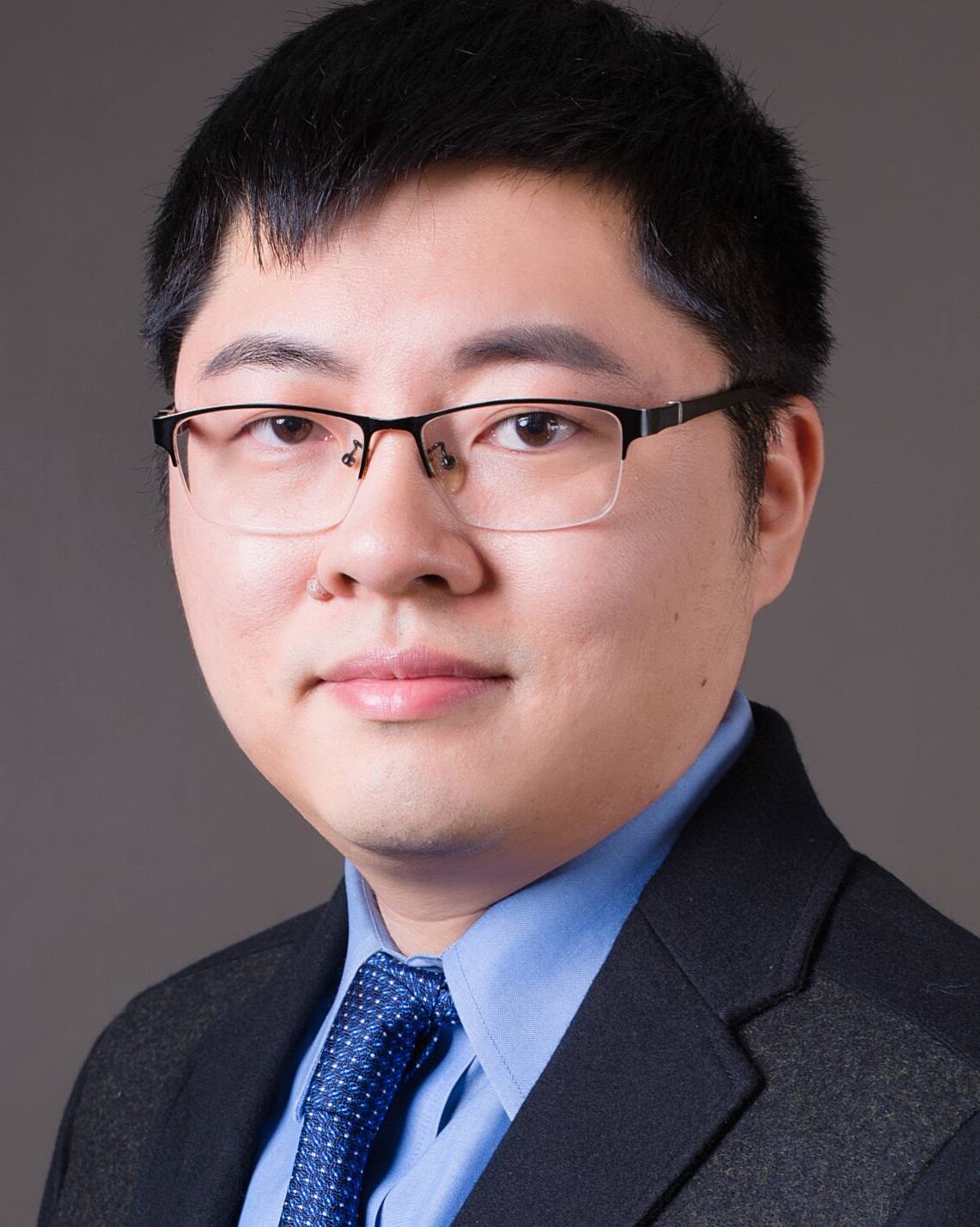}}]{Ming Li}(Senior Member, IEEE)
received his Ph.D. in Electrical Engineering from University of Southern California in 2013. He is currently an Associate Professor of Electrical and Computer Engineering at Duke Kunshan University. He is also an Adjunct Professor at School of Computer Science in Wuhan University. His research interests are in the areas of audio, speech and language processing as well as multimodal behavior signal processing. He has published more than 180 papers and served as the member of IEEE speech and language technical committee, APSIPA speech and language processing technical committee. He is an area chair at Interspeech 2016, 2018, 2020 and 2024, as well as the technical program co-chair of Odyssey 2022 and ASRU 2023. Works co-authored with his colleagues have won first prize awards at Interspeech Computational Paralinguistic Challenges 2011, 2012 and 2019, ASRU 2019 MGB-5 ADI Challenge, Interspeech 2020 and 2021 Fearless Steps Challenges, VoxSRC 2021, 2022 and 2023 Challenges, ICASSP 2022 M2MeT Challenge, IJCAI 2023 ADD challenge and ICME 2024 ChatCLR challenge. He received the IBM faculty award in 2016, the ISCA Computer Speech and Language 5-years best journal paper award in 2018 and the youth achievement award of outstanding scientific research achievements of Chinese higher education in 2020. He is a senior member of IEEE.
\end{IEEEbiography}

\end{document}